\documentclass[
  a4paper,
  preprint,
  5p,
  twocolumn,
  authoryear,
  ]{elsarticle}

\usepackage[utf8]{inputenc}

\usepackage[british]{babel}
\usepackage{newtxtext}  
\usepackage{newtxmath}
\usepackage[unicode=true,hidelinks]{hyperref}
\usepackage{microtype}
\usepackage[svgnames]{xcolor}

\usepackage{mathtools}
\usepackage{physics}
\usepackage{siunitx}

\usepackage{xifthen}  

\usepackage{ccicons}  

\newcommand*{\Rey}{\mathit{Re}}

\newcommand*{\Lmean}[1]{{\langle{#1}\rangle}}  
\newcommand*{\DoVec}{\vb{D}_0}  
\newcommand*{\DeltaV}{\delta\vb{v}}
\newcommand*{\DeltaA}{\delta\vb{a}}
\newcommand*{\Svv}{\Lmean{\DeltaV_0^2}}
\newcommand*{\Sva}{\Lmean{\DeltaV_0 \cdot \DeltaA_0}}
\newcommand*{\Rsq}[1][]{\Lmean{{\vb{R}#1}^2}}  
\newcommand*{\Rsqt}{\Rsq[(t)]}                 
\newcommand*{\Dsq}[1][]{\Lmean{{\vb{D}#1}^2}}  

\newcommand*{\Sau}{S_{au}}

\newcommand*{\RsqPrime}[1][]{\Rsq[^{\prime}#1]}  
\newcommand*{\RsqtPrime}{\RsqPrime[(t)]}         
\newcommand*{\SvvPrime}{\Lmean{\DeltaV_0^{\prime 2}}}
\newcommand*{\SvaPrime}{\Lmean{\DeltaV_0' \cdot \DeltaA_0'}}

\newcommand*{\Dij}[1][ij]{\Delta_{#1}}
\newcommand*{\Dxx}{\Dij[xx]}
\newcommand*{\Dyy}{\Dij[yy]}
\newcommand*{\Dzz}{\Dij[zz]}
\newcommand*{\Dxy}{\Dij[xy]}

\begin{document}

\title{%
  Relative dispersion of particle pairs in turbulent channel flow\tnoteref{cc}
}

\tnotetext[cc]{%
  \ccbyncnd{}
  This work is licensed under a Creative Commons CC BY-NC-ND 4.0 license
  (\url{https://creativecommons.org/licenses/by-nc-nd/4.0/})
}

\author[LMFA]{J.I.~Polanco\corref{corr}}
\ead{juan-ignacio.polanco@univ-lyon1.fr}

\author[LMFA]{I.~Vinkovic}
\ead{ivana.vinkovic@univ-lyon1.fr}

\author[LEGI]{N.~Stelzenmuller}

\author[LEGI]{N.~Mordant}

\author[ENSLyon]{M.~Bourgoin}

\cortext[corr]{Corresponding author}

\address[LMFA]{%
  Laboratoire de Mécanique des Fluides et d’Acoustique, UMR 5509,
  Ecole Centrale de Lyon, CNRS,\\
  Université Claude Bernard Lyon 1, INSA Lyon, 36 av. Guy de Collongue, F-69134
  Ecully, France
}

\address[LEGI]{%
  Laboratoire des Ecoulements Géophysiques et Industriels,
  Université Grenoble Alpes \& CNRS, \\
  Domaine Universitaire, CS 40700, F-38058 Grenoble, France
}

\address[ENSLyon]{%
  Laboratoire de Physique, École Normale Supérieure de Lyon, Université de Lyon,
  \\
  CNRS, 46 Allée d’Italie, F-69364 Lyon, France
}

\begin{abstract}

Lagrangian tracking of particle pairs is of fundamental interest in a large
number of environmental applications dealing with contaminant dispersion
and passive scalar mixing.
The aim of the present study is to extend the observations available in
the literature on relative dispersion of fluid particle pairs to
wall-bounded turbulent flows, by means of particle pair tracking in direct numerical
simulations (DNS) of a turbulent channel flow.
The mean-square change of separation between particle pairs follows a clear
ballistic regime at short times for all wall distances.
The Eulerian structure functions governing this short-time separation are
characterised in the channel, and allow to define a characteristic time scale
for the ballistic regime, as well as a suitable normalisation of the mean-square
separation leading to an overall collapse for different wall distances.
Through fluid particle pair tracking backwards and forwards in time, the
temporal asymmetry of relative dispersion is illustrated.
At short times, this asymmetry is linked to the irreversibility of turbulence,
as in previous studies on homogeneous isotropic flows.
The influence of the initial separation (distance and orientation) as well as
the influence of mean shear are addressed.
By decomposing the mean-square separation into the dispersion by the fluctuating
velocity field and by the average velocity, it is shown that the influence of
mean shear becomes important at early stages of dispersion close to the wall
but also near the channel centre.
The relative dispersion tensor $\Dij$ is also presented and particularly the sign and
time evolution of the cross-term $\Dxy$ are discussed.
Finally, a ballistic cascade model previously proposed for homogeneous
isotropic turbulence is adapted here to turbulent channel flows.
Preliminary results are given and compared to the DNS\@.
Future developments and assumptions in two particle stochastic models can be
gauged against the issues and results discussed in the present study.

\end{abstract}


\maketitle
\bibliographystyle{elsarticle-harv}

\section{Introduction}

The transport and mixing of passive components by turbulent flows are commonly
encountered in a large number of environmental and industrial applications.
Many atmospheric pollution studies have investigated contaminant
dispersion in the context of Lagrangian tracking of single particles
\citep{Hoffmann2016, Fung2005a, Angevine2013}.
In these studies, mesoscale meteorological models are often coupled with a
Lagrangian particle dispersion model providing a numerical method for simulating
the dispersion of passive pollutants in the atmosphere by means of a large
ensemble of Lagrangian particles moving with the modelled flow velocity field
\citep{Fung2005a}.
Recently, relative displacement measurements from balloons and drifters have
been conducted both in the atmosphere and the ocean
\citep{Lumpkin2010, LaCasce2010, Koszalka2009}.

Relative pair dispersion is of theoretical interest because particle pairs
simultaneously sample the velocity field at different positions.
In fluid flows, the short-term mean-square difference between tracer particle
velocities is equivalent to the second-order Eulerian velocity structure function, which
is related to the turbulent kinetic energy spectrum.
The turbulent kinetic energy at a particular scale determines how a tracer cloud
is stirred relative to its centre of mass.

In the last few decades, advances in experimental techniques and computational
power have enabled the characterisation of particle trajectories and relative
pair dispersion in canonical turbulent flows.
Most of these studies have dealt with fluid tracers in homogeneous isotropic
turbulence (HIT) and have led to a greatly increased understanding of the
Lagrangian properties of turbulent flows \citep{toschi_lagrangian_2009}, and in
particular, of the mechanisms of transport and diffusion of tracer particles in
isotropic flows.
A review on recent advances in experiments, direct numerical simulations (DNS)
and theoretical studies on particle pair dispersion has been provided by
\citet{salazar_two-particle_2009}, additionally to the review of
\citet{Sawford2001} on two-particle Lagrangian stochastic models.
As described by \citet{Sawford2001}, particle pair Lagrangian stochastic models
are a suitable tool for predicting dispersion of contaminant plumes in
turbulence, since separation statistics of particle pairs can be directly
related to the concentration covariance and to the dissipation of scalar
fluctuations.

Fewer studies have dealt with particle pair dispersion in anisotropic and
inhomogeneous turbulence, which is however ubiquitous in atmospheric flows.
In particular, most real flows are characterised by the presence of mean shear
or solid boundaries, which effectively suppress the movement of the fluid in the
wall normal direction.
This results in turbulent flows that are anisotropic due to the presence of a
mean shear, and inhomogeneous because of confinement by the walls.
Near the walls, turbulent fluctuations are mainly described by the formation of
large-scale organised structures that are elongated in the mean flow direction
\citep{smits_highreynolds_2011, Stanislas2017}.

As first described by \citet{richardson_atmospheric_1926}, turbulence can
greatly enhance the pair separation process.
In his seminal paper, Richardson proposed that the separation of two tracers in
a turbulent flow can be described (in a statistical sense) by a diffusive
process, with a non-constant diffusion coefficient $K(D)$ which depends on the
separation $D$ between the two particles.
When $D$ is within the inertial subrange of a turbulent flow (that is, much
larger than the dissipative scale $\eta$ and much smaller than the scale of the
largest turbulent eddies $L$), Richardson found from measurements that
the diffusion coefficient $K(D)$ is proportional to $D^{4/3}$, which is since
known as Richardson's $4/3$ law.
As later shown by \citet{Obukhov1941}, the same relation can be derived from
dimensional arguments in the framework of K41 local isotropy theory
\citep{Kolmogorov1941a}.
This requires the additional hypothesis that there is a loss of memory of the
initial condition, such that the initial pair separation $D_0$ no longer plays a
role in the separation process \citep{Batchelor1950}.
As a consequence, the mean-square separation between two particles is expected
to grow as $\Lmean{D^2(t)} = g \varepsilon t^3$ when $D$ is in the inertial
subrange, where $\varepsilon$ is the mean turbulent energy dissipation rate, and
the non-dimensional coefficient $g$, known as Richardson's constant, is expected
to have an universal value.

As mentioned above, the initial separation $D_0$ must be taken into account
at short separation times \citep{Batchelor1950}.
This dependency can be expressed as a short-term ballistic growth of the mean-square
separation:
\begin{equation}
  \Rsqt =
  \Lmean{{(\vb{D}(t) - \DoVec)}^2} = \Lmean{\DeltaV_0^2} t^2
  \quad
  \text{for }
  t \ll t_B,
  \label{eq:R2_short_time_intro}
\end{equation}
where $\vb{D}(t)$ is the instantaneous particle separation vector and $\DoVec
= \vb{D}(0)$, $\DeltaV_0$ is the initial relative velocity between the
particles, and $t_B$ is a characteristic time scale of the
ballistic regime, that may be related to the characteristic time scales of the
turbulent flow.
Equation~\eqref{eq:R2_short_time_intro} can be obtained from the Taylor
expansion of $\vb{D}(t)$ about $t = 0$.
The average $\Lmean{\cdot}$ is taken over an ensemble of particle pairs
initially separated by $\DoVec$.
In HIT, if $D_0 = \abs{\DoVec}$ is within the inertial
subrange, the ballistic time $t_B$ may be taken as proportional to the
eddy-turnover time at the scale $D_0$ (also referred to as the Batchelor time
scale), i.e.\ $t_E = D_0^{2/3} \varepsilon^{-1/3}$ \citep{Batchelor1950}.

By following two million passive tracers in a direct numerical simulation,
\citet{biferale_lagrangian_2005} found high levels of intermittency for travel
times up to ten Kolmogorov time scales in pair dispersion statistics in HIT at
$\Rey_\lambda=260$.
The authors proposed an alternative method for calculating Richardson's constant
by computing statistics at fixed separations.
Also in HIT, \citet{Rast2011} studied pair dispersion by analysing the
time scale $t_B$ during which particle pairs remain together before the
separation increases significantly in a simplified point-vortex flow model.
The authors suggested that pair separation may be understood as an average
over separations which follow Richardson's scaling but each over a fluctuating
time delay $t_B$.

Relative dispersion in HIT is known to be a time-asymmetric
process.
That is, when fluid particles are tracked backwards in time (starting from an
imposed \emph{final} separation),
they tend to separate faster than in the
forward case \citep{Sawford2005, berg_backwards_2006, Buaria2015}.
Recently, \citet{Jucha2014} and \citet{bragg_forward_2016} linked this temporal
asymmetry at short times to
the irreversibility of turbulence, which can be understood as the directionality
of the turbulent energy cascade (from large to small scales in 3D turbulence).
Moreover, \citet{bragg_forward_2016} compared backward and forward in time
dispersion statistics for inertial particle pairs.
They found that the ratio of backwards to forwards in time mean-square
separation may be up to an order of magnitude larger for inertial particles than
for fluid particles in isotropic turbulence.
Inertial particles were found to experience an additional source of
irreversibility arising from the non-local contribution of their velocity
dynamics.

Richardson's super-diffusive regime described above requires the existence
of an intermediate time range in which the following two conditions are
simultaneously satisfied: (1) the initial separation has been
forgotten ($t \gg t_B$), and (2) particle separation remains small enough such
that their trajectories are still correlated ($D(t) \ll L$).
The second condition is equivalent to $t \ll T_L$, where $T_L$ is the Lagrangian
integral time scale \citep{salazar_two-particle_2009}.
This implies large scale separation which occurs for turbulent flows at very
high Reynolds numbers.

In inhomogeneous and anisotropic turbulent flows, the relative dispersion problem is more complex,
since the statistics depend not only on the magnitude, but also on the direction
of the initial particle separation vector $\DoVec$ and on the initial particle
position.
Moreover, particles do not separate equally in each direction.
Therefore, the mean-square separation $\Rsqt$ can be generalised
into a dispersion tensor $\Delta_{ij}(t) = \Lmean{R_i(t) R_j(t)}$
\citep{Batchelor1952} containing more than a single independent component (as
opposed to the isotropic case).

The case of a homogeneous shear flow was studied by direct numerical
simulations (DNS) by \citet{Shen1997}.
The authors observed that particles separate faster when they are initially oriented in
the cross-stream direction, that is, when they are in regions of different
streamwise mean velocities.
Moreover, regardless of their initial separation vector, over time their
mean-square separation becomes larger in the streamwise direction than in the
spanwise and cross-stream directions.
\Citet{Celani2005} studied the competition between the effects of turbulence
fluctuations and a linear mean shear on particle separation using a simple
analytical model.
They proposed the existence of a temporal transition between a first stage of
separation, where turbulent fluctuations dominate and Richardson's law can be
expected to hold, and a second stage where mean shear becomes dominant.
The transition is expected to happen at a crossover time which is proportional
to the characteristic time scale of the mean shear.

More recently, \citet{Pitton2012} studied the separation of inertial particle pairs in a
turbulent channel flow using DNS at a friction Reynolds number $\Rey_\tau = 150$.
Results for inertial particles were compared to fluid tracers.
The authors observed that mean shear induces a super-diffusive regime at large
times, when particle separation becomes of the order of the largest scales
of the flow.
Arguably due to an insufficient separation of scales, Richardson's regime was
not clearly identified.
\citet{Pitton2012} removed the effect of mean shear by tracking particles which
follow the fluctuating velocity field.
They found that, although pair separation is importantly reduced at long times
compared to the case with mean shear, separation in the streamwise direction
remains dominant over the wall-normal and spanwise separations.

The DNS of \citet{Pitton2012} revealed the fundamental role played by inertial
particle-turbulence interactions at small scales in the initial stages of pair
separation.
The authors found a super-diffusive pair dispersion of inertial particles in
channel flow.
This super-diffusion at short times exhibited strong dependency on particle
inertia, and persisted even when the influence of mean shear was removed by the
procedure described in the previous paragraph.
Using DNS and Lagrangian tracking of inertial particles, \citet{Sardina2012}
studied turbulence-induced wall accumulation of inertial particles
(turbophoresis) and small-scale clustering.
The authors showed that for inertial particles, the clustering intensity in the
near-wall region is directly correlated with the strength of the turbophoretic
drift.
In the case of inertial particles, clustering and near-wall accumulation are
expected to strongly influence particle pair dispersion statistics.
\Citet{Lashgari2016} used a DNS coupled with immersed boundary methods to study
the collision kernel and relative pair statistics of finite-size solid particles
in turbulent channel flows for a wide range of volume fractions and Reynolds
numbers.
The authors found that the particle relative velocity and clustering are clearly
influenced by inertia and particle concentration.
Recently, \citet{Fornari2018} studied polydispersed particle pair statistics
also by DNS and an immersed boundary method accounting for finite-size effects.
The radial distribution function and the average normal relative velocity
between two approaching particles were computed in order to estimate the
collision kernel.
Collision statistics were found to be dominated by the behaviour of smaller
particles.
\Citet{Fornari2018} calculated that on average inertial particles stay during
$t = 2.5 h/U_0$ within a radial distance of one particle radius, indicating that
for polydispersed inertial particles long times are needed before a particle
pair breaks.

The aim of this study is to extend the available literature on relative
dispersion of fluid particle pairs in anisotropic and inhomogeneous turbulent
flows.
Pair dispersion statistics are obtained here by DNS in a turbulent channel flow
at a Reynolds number based on the friction velocity, $\Rey_\tau \approx 1440$.
Particle pairs are tracked backwards and forwards in time to characterise the
time asymmetry of relative dispersion.
New results show that a simple ballistic dynamics accurately reproduces the
initial pair separation regime and the short-term temporal asymmetry of
relative dispersion, regardless of the local turbulence properties (anisotropy,
turbulent structures) in the channel.
Based on this description,
time scales relevant to relative dispersion are discussed and a
suitable normalisation is proposed, leading to similarity of mean-square
separation statistics for initial wall distances $y^+_0 \gtrsim 60$.
The influence of mean shear and of the initial distance and orientation of the
particle pair separation are addressed in this work.
Finally, a simple ballistic cascade model
accounting for the influence of mean
shear is presented.
The model reproduces the main features of the initial stages of dispersion as
observed by DNS\@.
Suitable data and results are provided with which the assumptions and
predictions of two-particle stochastic models \citep{Sawford2001} can be tested.
Particularly, the direct numerical simulation results discussed here may give
more detailed information against which future developments and modelling
assumptions can be gauged.

The structure of the paper is the following.
We first present the numerical approach (Section~\ref{sec:numerical_approach}).
In Section~\ref{sec:mean_square_separation} the mean-square separation evolution
in time is discussed.
A characteristic time scale of the ballistic regime and a normalisation of the
relative dispersion at different wall distances are proposed.
The influence of mean shear is addressed in Section~\ref{sec:mean_shear}.
The analysis of the relative dispersion tensor is given in
Section~\ref{sec:dispersion_tensor}.
Finally, an adaptation of the ballistic cascade model initially proposed by
\citet{Bourgoin2015} to the case of a turbulent channel flow is presented in
Section~\ref{sec:ballistic_dispersion_model} together with preliminary results
and comparisons with DNS\@.
Section~\ref{sec:conclusion} is devoted to the conclusion.

\section{Numerical approach}%
\label{sec:numerical_approach}

\begin{figure}[tb]
  \centering
  \includegraphics[]{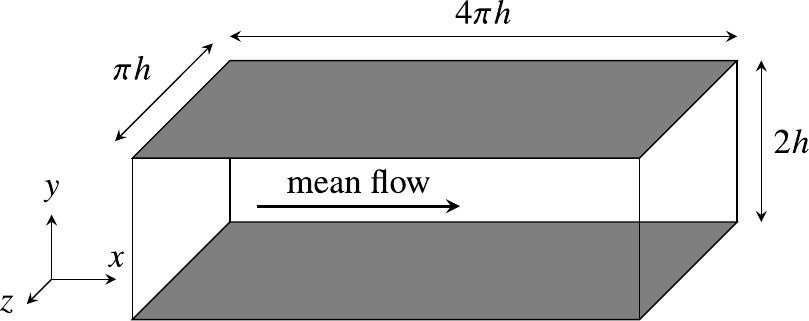}
  \caption{%
    Channel dimensions and coordinate system.
  }\label{fig:channel}
\end{figure}

We perform direct numerical simulations to study the relative dispersion
of fluid particles in a turbulent channel flow between two parallel walls
separated by a distance $2h$, as illustrated in Fig.~\ref{fig:channel}.
The Reynolds number based on the mean velocity $U_0$ at the channel centre is
$\Rey = U_0 h / \nu = \num{34000}$, where $\nu$ is the kinematic viscosity of
the fluid.
This corresponds to a friction Reynolds number
$\Rey_\tau = u_\tau h / \nu \approx \num{1440}$, where
$u_\tau = {(\tau_w/\rho)}^{1/2}$ is the friction velocity associated to the mean
shear at the walls $\tau_w$.
In the following, the superscript $+$ is used to indicate physical quantities
normalised by $u_\tau$ and $\nu$.

In the DNS, the Navier-Stokes equations are solved using a pseudo-spectral
method \citep{buffat_efficient_2011}.
The solver is coupled with Lagrangian tracking of fluid particles.
The numerical domain is periodic in the streamwise ($x$) and the spanwise ($z$)
directions, where the solution is decomposed into Fourier modes.
In the wall-normal ($y$) direction, a Chebyshev expansion is applied and no-slip
boundary conditions are enforced at the channel walls.
As in \citet{Stelzenmuller2017}, the domain size is $L_x \times L_y \times L_z =
4\pi h \times 2h \times \pi h$, and the velocity field is decomposed into
$\num{2048 x 433 x 1024}$ modes.
In physical space, this corresponds to a grid spacing $\Delta x^+ = 8.9$ and
$\Delta z^+ = 4.4$ in the periodic directions, while the wall-normal spacing
$\Delta y^+$ ranges from 0.04 at the wall to 10.5 at the channel centre.
The Eulerian velocity field $\vb{u}(\vb{x}, t)$ is advanced in time using an
explicit second-order Adams-Bashforth scheme with a time step
$\Delta t^+ = 0.033$.
The acceleration field is obtained in the Eulerian frame from the resolved
velocity according to
$\vb{a} =
\pdv{\vb{u}}{t} + \grad(\vb{u}^2/2) + (\grad \cross \vb{u}) \cross \vb{u}$.
Fluid particle tracking is achieved by interpolation of the velocity and
acceleration fields
at each particle position
using third-order Hermite polynomials.
Particles are advanced in space at each iteration
using the same
Adams-Bashforth scheme as for the Eulerian field.
Particle positions, velocities and accelerations are stored every 10
iterations (every $\Delta t^+_p = 0.33$).
The total simulation time is $T^+_\text{sim} \approx \num{1.7e4}$, or
equivalently $T_\text{sim} U_0 / h \approx 279$ based on the mean centreline
velocity $U_0$.

Dispersion statistics are obtained from two different sets of fluid particles,
labelled DS1 and DS2.
The dataset DS1 consists of \num{2e6} particles initialised at random positions
in the domain.
During post-processing, particle pairs are identified at chosen times
$t_0$ according to the criterion described further below, and relevant
statistics are computed over the temporal range $t \in [t_0 - T/2, t_0 + T/2]$.
This naturally allows to obtain backwards and forwards dispersion statistics,
and is similar to the approach described in \citet{berg_backwards_2006} and more
recently in \citet{Buaria2015}.
The temporal window length is chosen as $T^+ \approx \num{1.1e4}$, and the
spacing between two reference times $t_0$ is taken as
$\Delta t_0^+ \approx \num{1.3e3}$.

The criterion for particle pair identification in dataset DS1 is as follows.
Pairs separated by $|\DoVec| < D_0^{\text{max}}$ at $t_0$ are identified,
such that their centroids are located within bins of wall-normal distance
$y = y_0 \pm \delta y / 2$.
The maximum pair separation is taken as $D_0^{\text{max}} = 16\eta$,
where the
Kolmogorov length scale $\eta$, which varies with wall distance, is defined as
$\eta = {(\nu^3 / \varepsilon)}^{1/4}$.
Here, the mean turbulent energy dissipation rate is estimated as
$\varepsilon = \nu \overline{(\partial_j u'_i) (\partial_j u'_i)}$,
where $\vb{u}'(\vb{x}, t)$ is the instantaneous fluctuating velocity field.
The mean dissipation profile $\varepsilon(y)$ has been computed in the Eulerian
frame from the same DNS\@.
Pair dispersion statistics are computed over sets of particle pairs initialised
at the same reference wall distance $y_0$.
In wall units, the positions $y_0^+ =$ 20, 60, 200, 600 and 1000 are chosen (the
channel centre is at $y^+ = \num{1440}$).
The bin widths are taken as $\delta y = 8\eta$.
The Kolmogorov length scale ranges from $\eta^+ \approx 1.72$ at $y_0^+ = 20$ to
$\eta^+ \approx 5.31$ at $y_0^+ = 1000$.
Consequently, particles in the $y_0^+ = 20$ group may initially be located
within $0 \lesssim y^+ \lesssim 40$.
Due to the evolution of $\eta$ with wall distance, the total number of
identified particle pairs varies from roughly $\num{1.7e4}$ samples
at $y_0^+ = 20$, to $\num{1.5e6}$ samples at $y_0^+ = 1000$.
The dataset DS1 has already been used to study the acceleration of Lagrangian
tracers in a turbulent channel flow at the same Reynolds number
\citep{Stelzenmuller2017}.

Particles in dataset DS2 are initialised at chosen locations in order to
characterise the influence of the initial configuration of the pairs on
relative dispersion.
Each initial configuration is defined by 3 parameters: the initial wall distance
$y_0$ of one of the particles in the pair; the separation magnitude $D_0$
between the two particles; and the orientation of the pair separation $\vb{e}_0$, so that
their initial separation vector is $\DoVec = D_0 \vb{e}_0$.
In the simulations, we chose 10 initial wall distances $y_0^+$ ranging from $3$
to $1440$, combined with separations $D_0/\eta =$ 1, 4, 16 and 64, and
orientations in the three Cartesian directions ($\vb{e}_0 \in \{\vb{e}_x,
\vb{e}_y, \vb{e}_z\}$).
This results in 120 different initial configurations.
For each parameter combination, the size of the statistical sample (i.e.\ the
number of particle pairs) is roughly \num{20000}.
Only forward dispersion statistics are obtained from this dataset.
Applying the same approach to backward dispersion would require the storage of
an exceedingly large amount of Eulerian velocity fields, with a prohibitive cost
in terms of storage memory \citep{Sawford2005}.

\begin{figure}[tb]
  \centering
  \includegraphics[width=0.95\columnwidth]{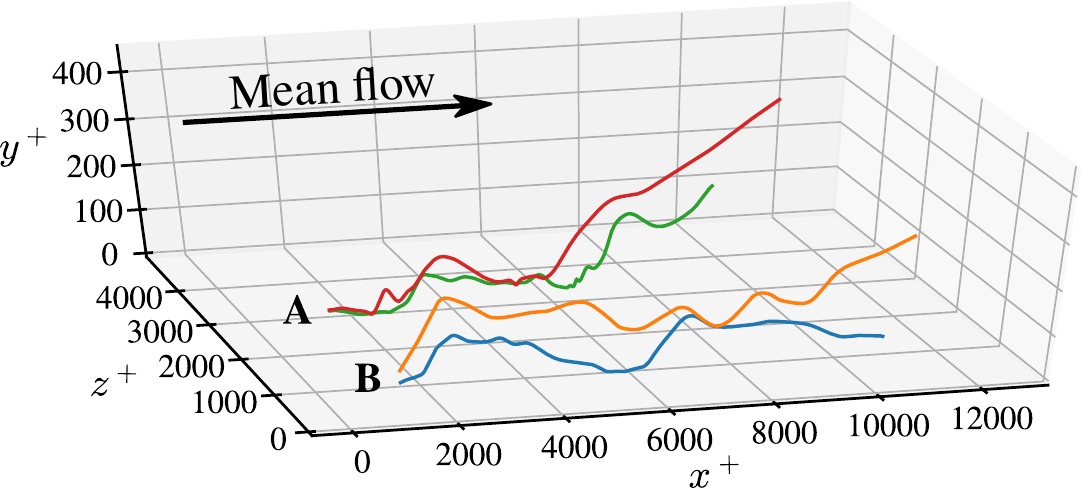}
  \caption{%
    Sample trajectories of two pairs of particles from dataset DS2.
    Trajectories are shown over $t^+ = \num{600}$.
    In both cases, the initial wall distance is $y_0^+ = \num{18}$ and the
    initial separation is $D_0 = \num{16}\eta$ ($D_0^+ = 27$).
    The pairs \textbf{A} and \textbf{B} are initially oriented in the spanwise
    ($z$) and wall-normal ($y$) directions, respectively.
  }\label{fig:pairs_3d}
\end{figure}

In Fig.~\ref{fig:pairs_3d}, the trajectories of two pairs of particles
initialised near the wall are shown.
At the initial time, both particle pairs differ only on the orientation of their
initial separation, with pair A being oriented in the spanwise direction,
and pair B in the wall-normal direction.
At the initial stage of separation, mean shear has no influence on the
separation of pair A, since both particles are at the same wall
distance $y^+$.
For relatively small wall-normal particle separations $D_y$, the influence of
turbulent fluctuations on separation statistics dominates over mean shear.
The case of pair B is different, since the two particles are initially in
regions of different mean velocity, and therefore shear effects are important
from the start.
As can be seen from Fig.~\ref{fig:pairs_3d}, under the influence of mean shear,
particles in pair B separate faster than in pair A following their release, reaching
larger separations $D$ at short times.
At larger times, the influence of the initial orientation is less noticeable, as
observed from comparable separations of pairs A and B at the end of all
trajectories ($t^+ = 600$).

\section{Mean-square separation}%
\label{sec:mean_square_separation}

\begin{figure}[tb]
  \centering
  \includegraphics[]{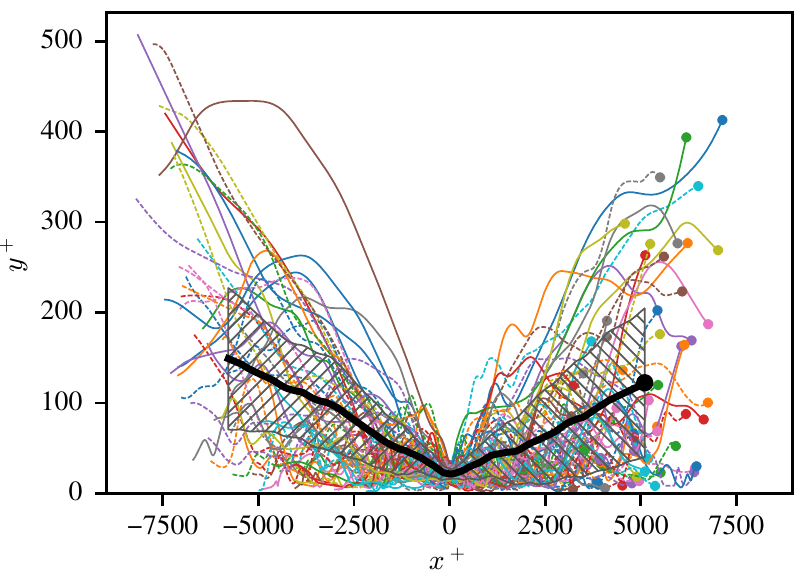}
  \caption{%
    Illustration of the Lagrangian averaging procedure.
    Thin curves represent trajectories of sample particle pairs with centroids
    located at $y^+ = y_0^+ \pm \delta y^+ / 2$ for a reference time $t = 0$
    (here $y_0^+ = 20$ and $\delta y^+ = 14$).
    The initial pair separation is $D_0 < 16\eta$ ($D_0^+ \lesssim 27$).
    Trajectories are shifted in the streamwise direction so that the particle
    pair centroids are at $x(t = 0) = 0$.
    The thick curve represents the Lagrangian average of the particle positions
    $\Lmean{\vb{x}(t)}$.
    The hatched area represents the standard deviation of the wall-normal
    particle pair separation $D_y(t)$, and is defined as the area between the
    curves $\Lmean{y(t)} \pm 0.5 \Lmean{D_y^2(t)}^{1/2}$.
    Trajectories are shown for time lags $t^+ \in [-400, 400]$.
  }\label{fig:lagrangian_mean_trajectories}
\end{figure}

Particle pair separation statistics are considered here in a fully-developed
wall-bounded turbulence by analysing the mean-square change of
separation between two particles, $\Rsqt =
\Lmean{{(\vb{D}(t) - \DoVec)}^2}$, where $\vb{D}(t)$ is the instantaneous
separation vector, and $\DoVec = \vb{D}(0)$ is the initial separation.
In HIT, statistics of $\vb{R}$ only depend on two
parameters: the initial particle separation distance $D_0 = \abs{\DoVec}$ and
time $t$.
In channel flow, as a consequence of anisotropy and inhomogeneity, such
statistics also depend on the initial orientation of the pair $\vb{e}_0$ (such that
$\DoVec = D_0 \vb{e}_0$) and on the initial wall-normal position
$y_0$ of one of the particles in the pair (such that the wall-normal position
of the other particle is $y_0 + \DoVec \cdot \vb{e}_y$).
Here, $\Lmean{\cdot}$ denotes an average over pairs of particles initially
located at the same $y_0$ and with the same initial orientation and separation
$\DoVec$.
In Fig.~\ref{fig:lagrangian_mean_trajectories}, this Lagrangian averaging
procedure is illustrated for sample particle pairs initially located at
$y_0^+ \approx 20$, and with initial separations $D_0 < 16\eta$.
The represented Lagrangian statistics are the mean particle position
$\Lmean{\vb{x}(t)}$ and the wall-normal mean-square separation $\Lmean{D_y^2(t)}$.

\begin{figure}[tb]
  \centering
  \includegraphics[]{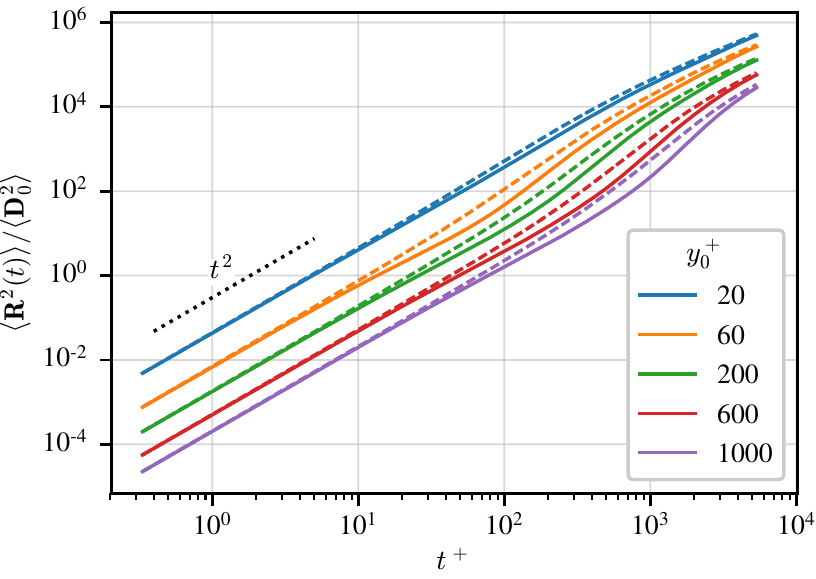}
  \caption{%
    Backward and forward mean-square change of separation $\Rsq$ normalised by the initial
    mean-square separation $\Lmean{\DoVec^2}$.
    Particle pairs are initially separated by $D_0 < 16\eta$ (dataset DS1).
    Different colours correspond to different initial wall distances $y_0^+$.
    Solid lines: forward dispersion.
    Dashed lines: backward dispersion.
  }\label{fig:R2_D0}
\end{figure}

Figure~\ref{fig:R2_D0} shows the time evolution of the mean-square change of separation
$\Rsq$ for initial separations $D_0 < 16\eta$, and for different initial
positions $y_0$ (dataset DS1).
In this case, statistics are averaged among all initial
separation vectors $\DoVec$ within a sphere of radius $16\eta$.
At short times, the ballistic regime predicted by
Eq.~\eqref{eq:R2_short_time_intro} is found for both backward and forward
dispersion, and for all wall distances.
Following this initial regime, a growing gap is observed at intermediate times
between backward and forward dispersion, with the former being
faster than the latter.
This is qualitatively consistent with observations in 3D HIT, described in the
introduction \citep{Sawford2005, berg_backwards_2006, Jucha2014,
bragg_forward_2016}.

In the following subsections, first the short-time ballistic dispersion regime
is analysed.
By considering the Taylor expansion of the separation at short times, the
influence of the second-order Eulerian velocity structure function and the
crossed velocity-acceleration structure function is emphasised.
The evolution of these structure functions is described in
Section~\ref{sec:structure_functions}.
Then, in Section~\ref{sec:ballistic_time_scales}, a suitable definition of the
ballistic time scale is presented, enabling the introduction of the normalised
mean-square separation in Section~\ref{sec:normalised_R2}.
The temporal asymmetry of pair dispersion statistics is then addressed in the
case of turbulent channel flow (Section~\ref{sec:temporal_asymmetry}), as well
as the influence of the initial separation distance and orientation
(Section~\ref{sec:initial_separation}).

\subsection{Short-time dispersion}%
\label{sec:short_time_dispersion}

To understand the observed short-time ballistic regime and the deviation that
follows, we consider the Taylor expansion of the separation between two
particles at short times, $\vb{D}(t) = \DoVec + \DeltaV_0 t + \frac{1}{2}
\DeltaA_0 t^2 + \order{t^3}$.
Here $\DeltaV_0$ and $\DeltaA_0$ are the relative particle velocity and
acceleration, respectively, at $t = 0$.
As a result, the short-time mean-square separation is expressed as
\begin{equation}
  \Rsq[](y_0, \DoVec, t) =
  \Svv t^2 + \Sva t^3 + \order*{t^4}
  \quad
  \text{for }
  t \ll t_B,
  \label{eq:R2_short_time_t3}
\end{equation}
where the characteristic time scale $t_B$ describes the duration of the
short-time regime.
At the leading order, the mean-square separation follows the ballistic regime
\citep{Batchelor1950}, during which particles travel at their initial
velocities.
The mean-square initial relative velocity $\Svv$ is equivalent to
the second-order Eulerian structure function
$S_2(\vb{x}_0, \DoVec) =
\overline{{\delta \vb{u}}^2}(\vb{x}_0, \DoVec) =
\overline{{[\vb{u}(\vb{x}_0 + \DoVec, t) - \vb{u}(\vb{x}_0, t)]}^2}$,
where $\vb{u}(\vb{x}, t)$ is the Eulerian velocity, and $\vb{x}_0$ is the
position of the first particle in the pair.
In channel flows, due to statistical homogeneity in the streamwise and spanwise
directions, the dependency of $S_2$ on $\vb{x}_0$ reduces to a dependency on the
wall-normal distance $y_0$.
In HIT, $S_2$ only depends on the separation $D_0 = \abs{\DoVec}$.
Moreover, when this separation is within the inertial subrange, K41 theory
predicts the well-known relation
$S_2(D_0) = \frac{11}{3} C_2 {(\varepsilon D_0)}^{2/3}$,
where $C_2$ is Kolmogorov's constant for the longitudinal second-order velocity
structure function, with $C_2 \approx \num{2.1}$
\citep{Sreenivasan1995, Pope2000}.

At the next order, the ballistic term in Eq.~\eqref{eq:R2_short_time_t3} is
corrected by a $t^3$ term whose coefficient $\Sva$ is equal to the crossed
velocity-acceleration structure function
$\Sau(\vb{x}_0, \DoVec) =
\overline{\delta \vb{u} \cdot \delta \vb{a}}(\vb{x}_0, \DoVec)$.
Under the conditions of local homogeneity and stationarity, if the spatial
increment $D_0$ is within the inertial subrange, the velocity-acceleration
structure function is given by
\begin{equation}
  \Sau(\vb{x}_0, \DoVec) =
  - (\varepsilon(\vb{x}_0) + \varepsilon(\vb{x}_0 + \DoVec)) =
  -2 \tilde{\varepsilon}(\vb{x}_0, \DoVec),
  \label{eq:S_au_epsilon}
\end{equation}
where $\tilde{\varepsilon}$ is the turbulent dissipation rate averaged among the
two probed positions \citep{Mann1999, Hill2006}.
This relation is exact in the limit of infinite Reynolds numbers, and is the
Lagrangian equivalent of Kolmogorov's 4/5 law \citep{Frisch1995}.
The negative sign of $\Sau$ is associated with the direction of the turbulent
cascade, from large to small scales in 3D turbulence.
Thus, under the assumptions for Eq.~\eqref{eq:S_au_epsilon}, the $t^3$ term of
Eq.~\eqref{eq:R2_short_time_t3} is negative for forward dispersion ($t > 0$)
and positive for backward dispersion ($t < 0$).
This explains the short-time temporal asymmetry of relative dispersion in
isotropic flows \citep{Jucha2014}.

\subsection{Structure functions
  \texorpdfstring{$S_2$}{S2} and \texorpdfstring{$\Sau$}{Sau}
}\label{sec:structure_functions}

The evolution of the structure functions introduced above with wall distance and
with spatial increment is investigated in order to describe the short-time
dispersion regime given by Eq.~\eqref{eq:R2_short_time_t3}.
To our knowledge, few studies in literature have dealt with Eulerian structure
functions in wall-bounded turbulent flows.
Moreover, studies characterising
the crossed velocity-acceleration structure function $\Sau$ in such flows are
still lacking.
Existent works have focused on the logarithmic region of boundary layers,
and have mainly studied the second-order streamwise velocity
structure function $S_{xx}(y, \vb{D}) = \overline{\delta u_x^2}(y, \vb{D})$
for streamwise separations, $\vb{D} = D \vb{e}_x$
\citep[see e.g.][]{Davidson2014, Silva2015}.
Recently, \citet{Yang2017} proposed scalings for the complete fluctuating velocity
structure function tensor
$S'_{ij}(y, \vb{D}) = \overline{\delta u_i' \delta u_j'}(y, \vb{D})$ in the
logarithmic region, based on the attached-eddy model
\citep{Townsend1976}.
However, when considering spanwise separations ($\vb{D} = D \vb{e}_z$), their
predicted scalings do not match the results obtained from channel flow DNS at
moderate Reynolds number \citep{Lozano-Duran2014}.

\begin{figure}[htp]
  \centering
  \includegraphics{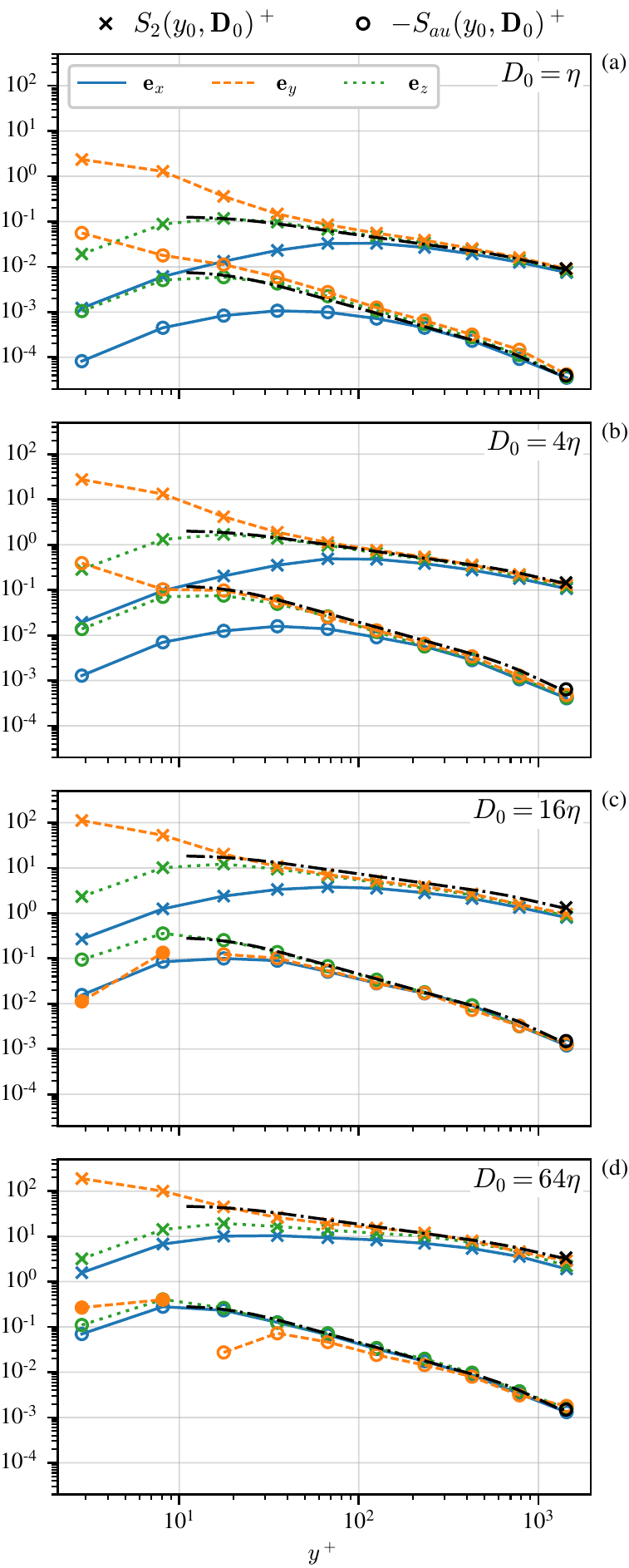}
  \caption{%
    Structure functions $S_2(y_0, \DoVec)$ (crosses) and
    $-\Sau(y_0, \DoVec)$ (circles) in wall units, for all initial
    configurations of dataset DS2.
    From top to bottom, $D_0/\eta = $ 1, 4, 16 and 64.
    Initial orientations are $\vb{e}_0 = \vb{e}_x$ (solid lines),
    $\vb{e}_y$ (dashed lines) and $\vb{e}_z$ (dotted lines).
    Filled circles represent positive values of $\Sau$.
    Black dash-dotted lines represent isotropic estimations of $S_2$ and
    $\Sau$.
    For small separations (subfigures a-b), a dissipation-range estimation is
    used,
    $S_2' = \varepsilon D_0^2 / (3\nu)$
    and
    $\Sau = \beta \varepsilon D_0^2 / (3\eta^2)$, with
    $\beta = \num{-0.16}$.
    For large separations (subfigures c-d), an inertial-range estimation is
    used,
    $S_2 = \frac{11}{3} C_2 {(\varepsilon D_0)}^{2/3}$ and
    $\Sau = -2 \varepsilon$, with $C_2 = \num{2.1}$.
  }\label{fig:structure_functions}
\end{figure}

We estimate $S_2$ and $\Sau$ across the channel from Lagrangian data
at $t = 0$ when particles of dataset DS2 are released.
The estimation is performed over all initial particle configurations,
namely for a range of wall distances $y_0$ and spatial displacement vectors
$\DoVec$.
The evolution with wall distance of the velocity and velocity-acceleration
structure functions, for different initial orientations and magnitudes of the
separation vector, is given in Fig.~\ref{fig:structure_functions}.
Since $\Sau$ is mostly negative (as expected in homogeneous
flows), we plot $-\Sau$.

In the near-wall region the structure functions display a strong
dependency on the orientation of the displacement $\DoVec$.
This anisotropy is due to wall confinement and the influence of mean shear.
The latter
only plays a role when the initial separation is in the wall-normal direction.
For this orientation, $S_2$ is expected to be larger since it includes a
contribution of the mean velocity increment
$\delta U = U(y + D_y) - U(y)$, where $U(y)$ is the mean streamwise
velocity across the channel.
This is confirmed by the curves of Fig.~\ref{fig:structure_functions}.
Moreover, near the wall $S_2$ is larger for spanwise than for streamwise
displacements, with a difference that is more pronounced for smaller separations
$D_0$.
This is due to the presence of streaks and
quasi-streamwise vortices, which induce a fluctuating velocity field that is
correlated for larger distances in the streamwise direction
\citep[see e.g.][]{Robinson1991}.
Hence, the velocity increment between two points in the near-wall region is
weaker if the points are aligned in the streamwise direction (since both
points are likely to be found within the same coherent structure), than in the
spanwise direction.

As shown in Fig.~\ref{fig:structure_functions}, the velocity-acceleration
structure function $\Sau$ is also anisotropic near the
wall.
For small separations ($D_0/\eta = $ 1 and 4), its behaviour is similar to that
of $S_2$, since its absolute value is larger for wall-normal displacements and
smaller for streamwise displacements.
As mentioned above, $\Sau$ is mostly negative.
Positive values are obtained in a few extreme cases when one of
the probed locations is at $y^+ < 10$ while the other is at $y^+ + D_{0y}^+$,
with $D_{0y} \geq 16\eta$ ($D_{0y}^+ \gtrsim 25$).
In these cases the velocity and acceleration increments describe the
relation between
the flow in the viscous sublayer (or the beginning of the buffer layer), and the beginning
of the logarithmic region.
Since these regions have very different dynamics, homogeneity is not expected to
hold on the resulting two-point statistics.
Furthermore, in these cases $\Sau$ is dominated by the scalar
product between the mean velocity and mean acceleration increments,
$\delta \vb{U} \cdot \delta \vb{A} = \delta U_x \delta A_x$.
In the buffer layer and the beginning of the logarithmic region, the mean streamwise
acceleration $A_x(y)$ is an increasing function of wall distance
\citep{yeo_near-wall_2010, Stelzenmuller2017}, similarly to the mean velocity
$U_x(y) = U(y)$.
This results in a positive product $\delta U_x \delta A_x$ when
locations across the buffer layer are sampled.

Away from the wall, the structure functions become nearly independent of the
displacement orientation, suggesting a return to isotropy towards the bulk of
the channel.
In general, this is observed for wall distances $y^+ \gtrsim 200$.
Still, a slight difference persists for $S_2$ at nearly all wall distances, with
the streamwise orientation resulting in a weaker structure function.
This may be associated with the persistence of very-large-scale motions in the
channel \citep[VLSMs;][]{smits_highreynolds_2011}.
A similar behaviour is observed for $\Sau$ at the smallest separations
$D_0/\eta =$ 1 and 4.

In Fig.~\ref{fig:structure_functions}, the $S_2$ profiles obtained from our DNS
at small separations $D_0/\eta = $ 1 and 4 are compared with the dissipation-range
estimation for the fluctuating part of the structure function,
$S_2' \sim \frac{1}{3} \overline{(\partial_j u_i') (\partial_j u_i')} D_0^2$,
which is derived from the first-order Taylor expansion
$\delta \vb{u}' \approx (\DoVec \cdot \nabla) \vb{u}'$ and the isotropy
assumption.
The estimation above can be expressed in terms of the mean turbulent dissipation
rate $\varepsilon = \nu \overline{(\partial_j u'_i) (\partial_j u'_i)}$.
For the two separations, the computed $S_2$ profiles closely match the
prediction, suggesting that separations up to $4\eta$ are not within the
inertial subrange.
Similarly, the $\Sau$ profiles at separations $\eta$ and $4\eta$ are compared to
the dissipation-range estimation
$\Sau \sim \frac{1}{3} \overline{(\partial_j u_i) (\partial_j a_i)} D_0^2$.
Dimensional considerations predict that
$\overline{(\partial_j u_i) (\partial_j a_i)} = \beta \varepsilon / \eta^2$,
with $\beta$ a non-dimensional constant.
The value $\beta = \num{-0.16}$ is found to fit the $\Sva$ data at
$D_0 = \eta$.
For $D_0 = 4\eta$, the prediction slightly overestimates the results obtained
from particle data in the bulk of the channel, hinting the beginning of the transition from
dissipation to the inertial regime.

Furthermore, we compare the larger separations $D_0/\eta =$ 16 and 64 with the
inertial-range K41 prediction for locally isotropic turbulence
$S_2(D_0) = \frac{11}{3} C_2 {(\varepsilon D_0)}^{2/3}$, where $\varepsilon$
varies with wall distance.
In the channel, the local isotropy condition may be expected to hold at
large-enough wall distances.
The obtained $S_2$ profiles accurately match the K41 prediction in the bulk of
the channel.
For spanwise separations, the estimation is accurate up to the
near-wall region.
Similarly, to verify the validity of relation~\eqref{eq:S_au_epsilon}, the
obtained $\Sau$ profiles at separations $16\eta$ and $64\eta$ are compared with
$-2\varepsilon(y)$.
For non-zero wall-normal displacements $D_y$, one has
$\varepsilon(y) \neq \tilde{\varepsilon}(y, D_y)$, so that the comparison is not
exactly equivalent to Eq.~\eqref{eq:S_au_epsilon} in the case of wall-normal
displacements.
Remarkably, the prediction holds almost exactly over a
wide range of wall distances.
This is especially true for spanwise displacements, for which a good
agreement is found at nearly all wall distances.

\subsection{Ballistic time scale}%
\label{sec:ballistic_time_scales}

The most suitable definition of the initial ballistic regime duration $t_B$ is
discussed in this section.
Originally, \citet{Batchelor1950} assumed this time as proportional to the
eddy-turnover time at scale $D_0$, i.e.\ $t_E = D_0^{2/3}
\varepsilon^{-1/3}$ \citep{Frisch1995}, when $D_0$ is in the inertial range.
An alternative is to consider the time at which the $t^2$ and $t^3$ terms in
Eq.~\eqref{eq:R2_short_time_t3} have the same magnitude, $t_0 = \Svv /
\abs{\Sva} = S_2(\vb{x}_0, \DoVec) / \abs{\Sau(\vb{x}_0, \DoVec)}$.
This characteristic time may be approximated by the dissipation- or
inertial-range predictions for the structure functions $S_2$ and $\Sau$
introduced in Section~\ref{sec:structure_functions}.
For separations $D_0$ in the dissipation range, this approximation is given by
$t_0^* = t_D = \tau_\eta / \beta$, where
$\tau_\eta = {(\nu / \tilde{\varepsilon})}^{1/2}$ is the Kolmogorov time scale.
For inertial-scale separations, the corresponding estimation is
$t_0^* = t_I = \frac{11}{6} C_2 D_0^{2/3} \tilde{\varepsilon}^{-1/3}$, which is
proportional to Batchelor's time scale.

\begin{figure}[!tb]
  \centering
  \includegraphics[]{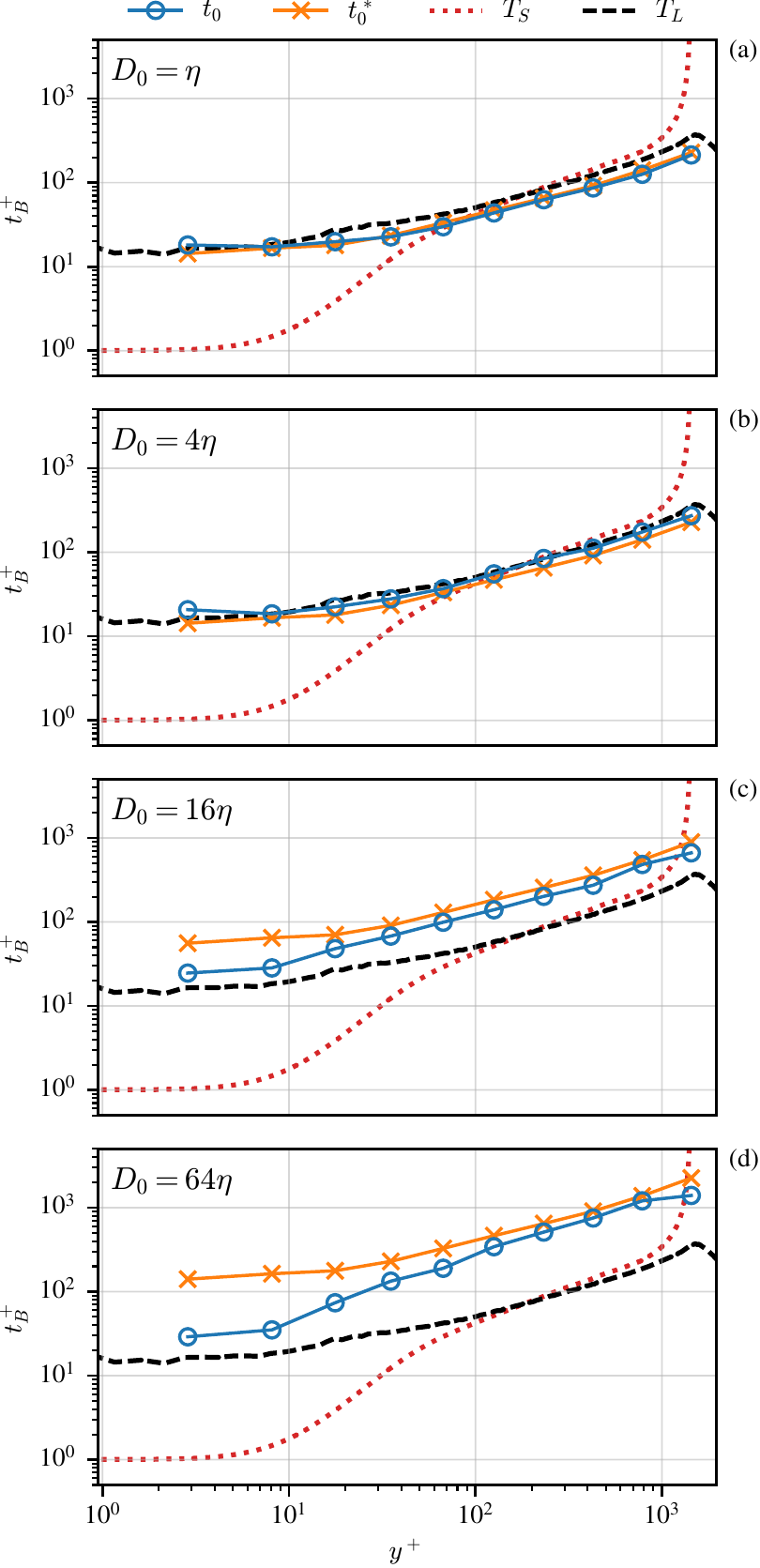}
  \caption{%
    Characteristic relative dispersion time scales in wall units along the
    channel width, for different initial separations $D_0$.
    From top to bottom, $D_0/\eta = $ 1, 4, 16 and 64.
    Results were obtained from dataset DS2.
    Pairs are initially oriented in the spanwise direction ($\vb{e}_0 =
    \vb{e}_z$).
    Circles,
    $t_0 = \Lmean{\DeltaV_0^2} /
    \abs{\Lmean{\DeltaV_0 \cdot \DeltaA_0}}$;
    crosses,
    $t_0^* = \tau_\eta / \beta$
    (subfigures a-b) or
    $t_0^* = \frac{11}{6} C_2 D_0^{2/3} \tilde{\varepsilon}^{-1/3}$
    (subfigures c-d).
    Non-dimensional constants are $C_2 = 2.1$ and $\beta = \num{-0.16}$.
    Also represented are the Lagrangian integral time scale $T_L$ (black
    dashed line) and the mean shear time scale $T_S = {(\dd U / \dd y)}^{-1}$
    (red dotted line).
  }\label{fig:timescales_yplus_z}
\end{figure}

The time scales $t_0$ and $t_0^*$ are computed for each of the initial
configurations of dataset DS2.
In Fig.~\ref{fig:timescales_yplus_z}, the results are shown for all sets of
particle pairs that were initially oriented in the spanwise direction.
For separations $D_0/\eta = $ 1 and 4, the dissipation-range form of $t_0^*$ is
plotted, while for $D_0/\eta = $ 16 and 64, the inertial-range approximation is
shown.
Also shown are the mean shear time scale across the channel, $T_S(y) =
{(\dd U(y) / \dd y)}^{-1}$ and the Lagrangian integral time scale $T_L(y)$, as
obtained in \citet{Stelzenmuller2017}.
Because of anisotropy, a different Lagrangian integral time
scale can be defined for each velocity component, $T_{L,i}$ for $i =$ 1, 2, 3
\citep{Stelzenmuller2017}.
Here, we take $T_L$ as the quadratic mean among the three velocity components,
$T_L^2 = \sum_i T_{L,i}^2 / 3$.

As shown in Fig.~\ref{fig:timescales_yplus_z}, the dissipation-range estimation
$t_0^* = \tau_\eta / \beta$ matches the ballistic time $t_0$ over all wall
distances for the smallest separation $D_0 = \eta$.
For $D_0 = 4\eta$, there is still good agreement between both time scales,
even though a weak departure from dissipation-range scaling is observed.
This departure is consistent with observations in
Section~\ref{sec:structure_functions} regarding the validity of the
dissipation-range estimation of $\Sau$ at $D_0 = 4\eta$.
The agreement between $t_0$ and $t_0^*$ shows the relevance of the
characteristic dissipation time $\tau_\eta$ on the ballistic separation
regime for small initial separations.
For separations $D_0 /\eta = $ 16 and 64, $t_0$ and the inertial-range
estimation $t_0^* = t_I$ mainly differ in the near wall region, and become
similar in the bulk of the channel.
As suggested by Fig.~\ref{fig:structure_functions}, the difference is explained
by a weakly overestimated inertial-range structure function $S_2$.

When compared to the Lagrangian integral time scale $T_L$, $t_0$ is of the same
order of magnitude for small separations, and considerably larger
than $T_L$ for larger separations.
This implies that scale separation is not achieved in this
channel flow and that an intermediate time range between $t_0$ and $T_L$ does not exist.
As a consequence, Richardson's super-diffusive regime cannot be observed under
the present flow conditions.

Finally, it is interesting to compare the ballistic time scale with the
characteristic time of the mean shear $T_S$.
As shown in Fig.~\ref{fig:timescales_yplus_z}, this time scale is small near the
wall, where shear is high, and grows far from the wall as shear decreases.
For separations $D_0/\eta =$ 1 and 4, $T_S$ is smaller than $t_0$ in the
near-wall region, up to $y^+ \approx 80$.
For larger separations, $T_S$ is smaller than $t_0$ everywhere in the
channel.
In these cases, mean shear is expected to influence relative dispersion
statistics since the beginning of particle pair separation.

\subsection{Normalised mean-square separation}%
\label{sec:normalised_R2}

\begin{figure}[tb]
  \centering
  \includegraphics[]{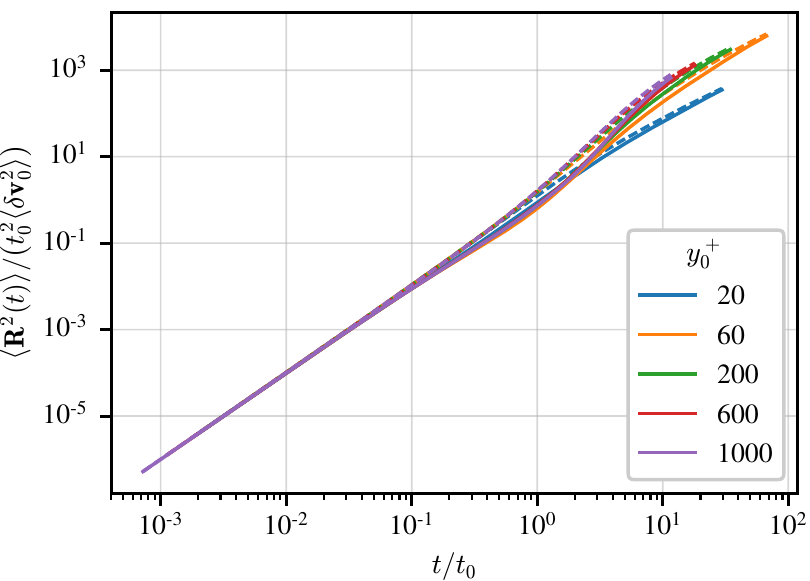}
  \caption{%
    Backward and forward mean-square separation normalised by the structure function
    $\Lmean{\DeltaV_0^2}$ and the characteristic ballistic time
    $t_0 = \Svv / \abs{\Sva}$.
    The initial pair separation is $D_0 < 16\eta$ (dataset DS1).
    Different colours correspond to different initial wall distances $y_0^+$.
    Solid lines: forward dispersion.
    Dashed lines: backward dispersion.
  }\label{fig:R2_ballistic}
\end{figure}

The time scale $t_0^*$ introduced in the previous section is
constructed from assumptions on the underlying turbulent flow, namely local
homogeneity and isotropy.
In contrast, $t_0$ is obtained according to purely kinematic considerations
(without any assumptions on the turbulent flow), and it is
chosen here as the characteristic ballistic time scale.
Thus, Eq.~\eqref{eq:R2_short_time_t3} can be rewritten as
\begin{equation}
  \frac{\Rsq[]}{\Svv t_0^2} =
  {\left(\frac{t}{t_0}\right)}^2 +
  s
  {\left(\frac{t}{t_0}\right)}^3 +
  \order*{t^4}
  \quad
  \text{for }
  t \ll t_0,
  \label{eq:R2_short_time_normalised}
\end{equation}
where $s \in \{-1, 1\}$ is the sign of $\Sva$.
In Fig.~\ref{fig:R2_ballistic}, dispersion curves of
Fig.~\ref{fig:R2_D0} are normalised by the expected ballistic regime according
to Eq.~\eqref{eq:R2_short_time_normalised}.
Under this scaling, forward dispersion curves associated to different wall distances collapse for
times up to $t \approx 2 t_0$, emphasising the relevance of the proposed
scaling.
At longer times, separation is accelerated for pairs that are
initially far from the wall.
A remarkable $t^2$ ballistic regime is observed for all wall distances.
Starting from $t \approx 0.1 t_0$, the mean-square separation deviates from the
initial ballistic regime becoming slightly slower for forward dispersion, and faster for
backward dispersion, consistently with a negative sign of
the $t^3$ term of Eq.~\eqref{eq:R2_short_time_normalised}.
Starting from $y_0^+ = 60$, normalised curves differ only slightly.
This is explained by the decay of inhomogeneity and anisotropy far from the
wall, resulting in Eulerian velocity and acceleration statistics which evolve
similarly with wall distance.
In the studied flow, $y^+ = 60$ is located at the beginning of the self-similar
logarithmic region \citep{Stelzenmuller2017}.

\begin{figure}[tb]
  \centering
  \includegraphics[]{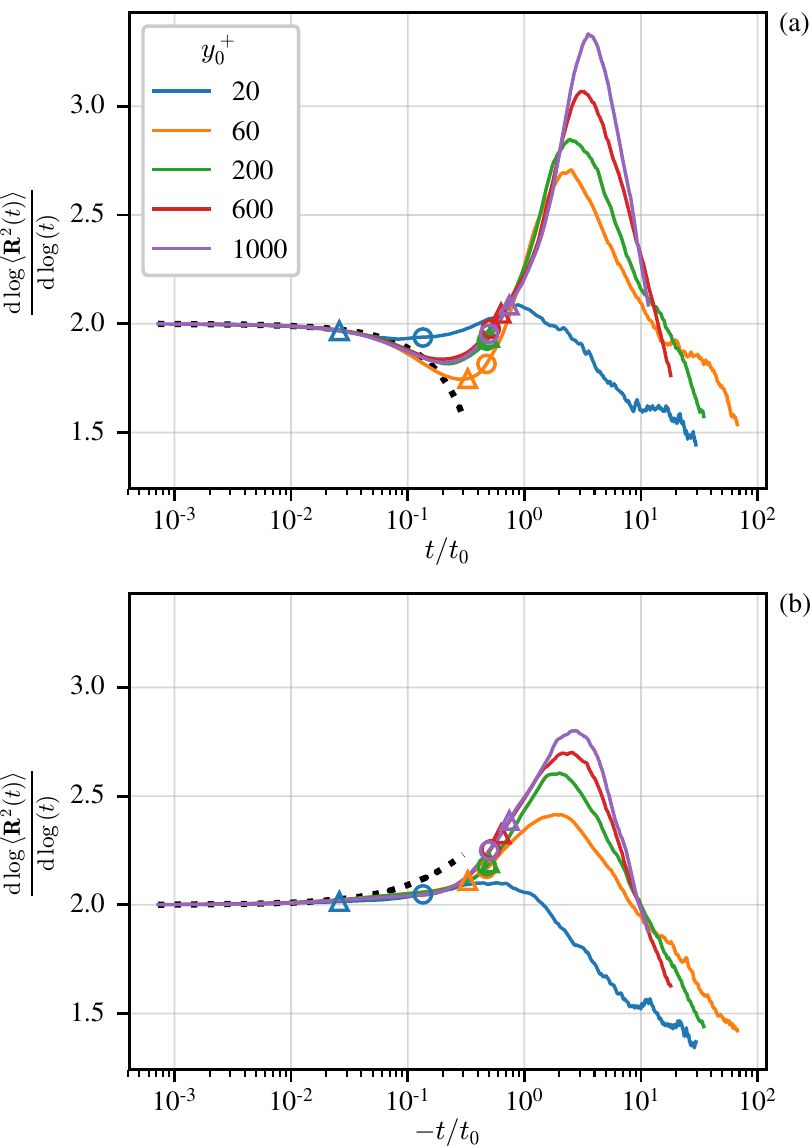}
  \caption{%
    Local scaling exponents of the mean-square separation process.
    (a) Forward dispersion.
    (b) Backward dispersion.
    The dotted line is derived from the truncated Taylor expansion
    $\Rsqt = \Svv t^2 + \Sva t^3$ with
    $\Sva < 0$.
    For each initial wall distance $y_0^+$, markers indicate the local value of
    the Lagrangian integral time scale $T_L$ (circles) and of the mean shear
    time scale $T_S$ (triangles).
  }\label{fig:R2_logdiff}
\end{figure}

Figure~\ref{fig:R2_logdiff} plots the local scaling exponent of the mean-square
separation, i.e., the local slope of the curves shown in
Fig.~\ref{fig:R2_ballistic}.
An initial plateau with a value of 2, corresponding to the ballistic regime, is
recovered both for forward and backward dispersion.
A deviation from this regime is observed as early as $|t|/t_0 \approx 0.01$, and
is given by a deceleration of the separation rate in the forward case,
and by an acceleration in the
backward case (as already seen in Fig.~\ref{fig:R2_ballistic}).
The early deviation from the ballistic regime can be associated to the purely
kinematic effect of the
$t^3$ term of Eq.~\eqref{eq:R2_short_time_normalised} when $s = -1$ (i.e.\
when $\Sva$ is negative).
This is confirmed by the comparison between the numerical results and the
truncated Taylor expansion of $\Rsqt$ in the figure.

At intermediate times, all cases present an increasing separation rate
that ends with a peak.
Except for the smallest initial wall distance $y_0^+ = 20$, the peak is found at
$2 < \abs{t}/t_0 < 5$.
The local scaling exponent reaches larger values in the forward case than in
the backward case.
A possible interpretation is that being faster, backward dispersion reaches
the normally-diffusive regime earlier than in the case of forward dispersion,
thus spending less time in the intermediate super-diffusive regime.
At long times, forward and backward separations match, consistently with
observations from Fig.~\ref{fig:R2_D0}.
In some cases, the local scaling exponents reach values around 3, which
is comparable to Richardson's $t^3$ super-diffusive regime.
However, Richardson's regime is not expected to be observed in this flow because
of the absence of scale separation and since mean shear is
important at early stages of dispersion (as discussed in
Section~\ref{sec:ballistic_time_scales}).
Moreover, as seen in Fig.~\ref{fig:R2_logdiff}, the peaks of the local scaling exponents
occur at times larger than the Lagrangian integral time scale $T_L$.
It may be argued that mean shear induces a super-diffusive regime at large
times.
As stated in the introduction, \citet{Pitton2012} also observed a shear induced
super diffusive regime for inertial particle separations of the order of the
largest flow scales.

\subsection{Temporal asymmetry}%
\label{sec:temporal_asymmetry}

\begin{figure}[tb]
  \centering
  \includegraphics[]{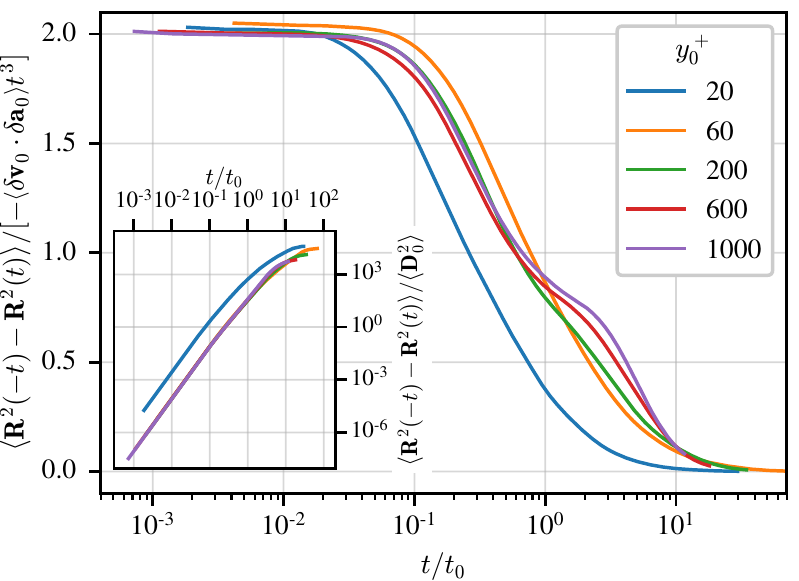}
  \caption{%
    Difference between backward and forward mean-square separation,
    compensated by $-\Sva t^3$.
    Particle pairs are initially separated by $D_0 < 16\eta$ (dataset DS1).
    Inset: mean-square separation difference compensated by the initial
    mean-square separation $\Lmean{\DoVec^2}$.
  }\label{fig:R2_time_asymmetry}
\end{figure}

The results discussed in Section~\ref{sec:normalised_R2} show the
temporal asymmetry of relative dispersion in turbulent channel flow.
At times following the initial ballistic separation regime, it has been
illustrated that backward dispersion is more effective than forward dispersion.
As suggested by \citet{Jucha2014}, the asymmetry at short times can be
explained by
subtracting the short-time expansion of the mean-square separation
(Eq.~\ref{eq:R2_short_time_t3}) for positive and negative times:
\begin{equation}
  \Rsq[(t)] - \Rsq[(-t)] =
  2 \Sva t^3 + \order*{t^5}
  \quad
  \text{for }
  t \ll t_0.
  \label{eq:R2_short_time_asymmetry}
\end{equation}
This difference is plotted in Fig.~\ref{fig:R2_time_asymmetry} compensated by
$\Sva t^3$.
As predicted by Eq.~\eqref{eq:R2_short_time_asymmetry}, a plateau with a value
of 2 is initially found for all initial wall distances.
A deviation from this plateau is observed starting from $t = 0.05 t_0$, which
is quantitatively consistent with results in HIT \citep{Jucha2014}.
This departure may be due to the neglected $t^5$ term in
Eq.~\eqref{eq:R2_short_time_asymmetry}, or to particle pairs sampling the
flow at scales larger than
$D_0$.
In the inset of Fig.~\ref{fig:R2_time_asymmetry}, the difference
$\Rsq[(-t)] - \Rsq[(t)]$ is normalised by the initial mean-square separation
$\Lmean{\DoVec^2}$.
The positive sign of this difference for all $y_0^+$ confirms that backward
dispersion evolves at faster rate than forward dispersion at all initial
positions.
Moreover, starting from $y_0^+ = 60$ similarity of the temporal asymmetry with
wall distance is observed.

\subsection{Influence of the initial separation}%
\label{sec:initial_separation}

The influence of the initial pair configuration, described by the
parameters $(y_0, \DoVec)$, on forward relative dispersion
is discussed here.
As described in Section~\ref{sec:short_time_dispersion}, short-time
dispersion follows a ballistic regime governed by the
second-order Eulerian velocity structure function $S_2(y_0, \DoVec)$,
described in Section~\ref{sec:structure_functions} for the set of
initial pair configurations studied in this work.

\begin{figure}[!tb]
  \centering
  \includegraphics[]{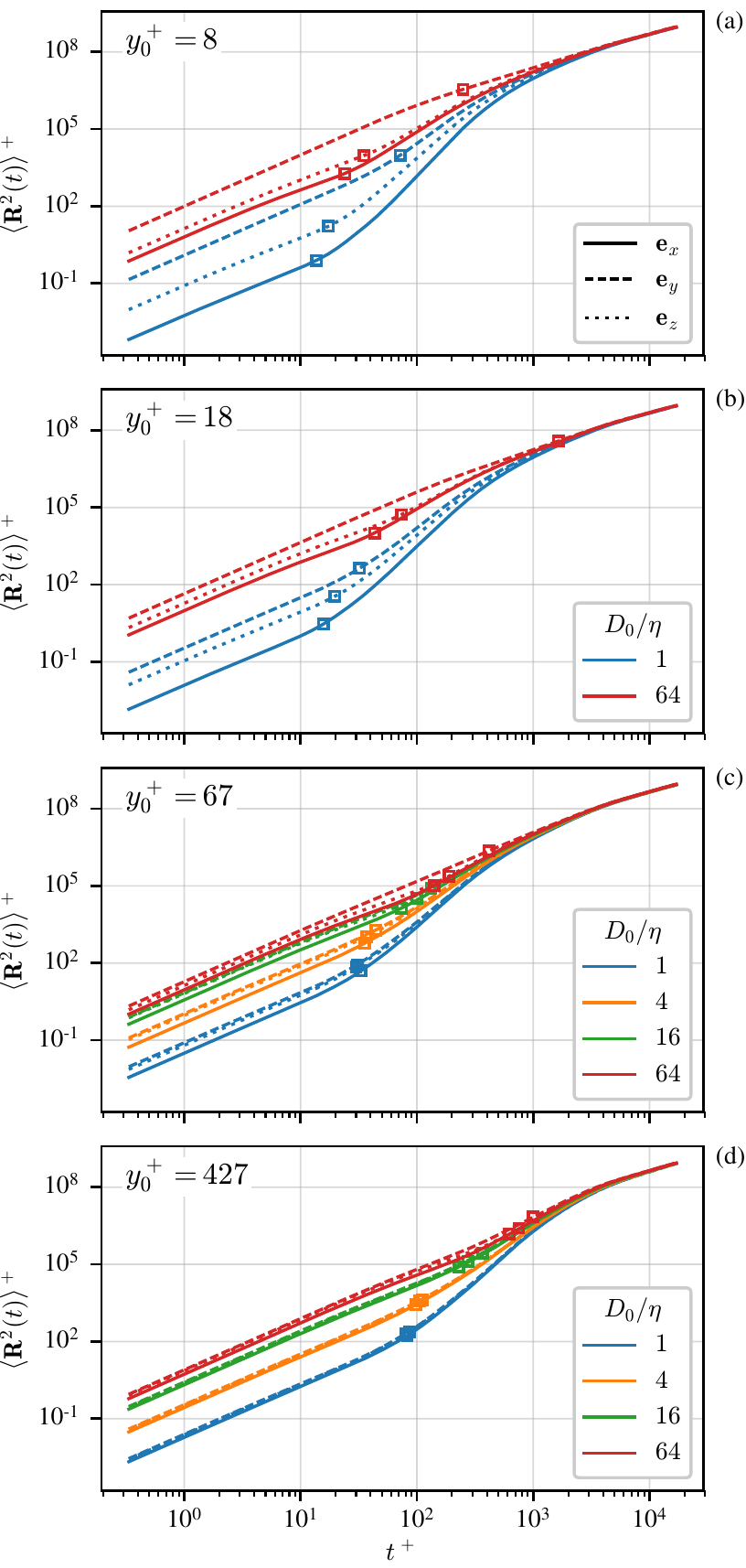}
  \caption{%
    Forward mean-square separation for different initial configurations, in wall units.
    Pairs are initially located at $y_0^+ = $ 8, 18, 67 and 427 (subfigures (a)
    to (d)).
    Line styles represent the initial orientation of the pairs:
    streamwise (solid lines), wall-normal (dashed lines) and spanwise (dotted
    lines).
    Line colours represent the initial separation $D_0/\eta$.
    Squares indicate the ballistic time $t_0$ associated to each initial condition.
    Results were obtained from dataset DS2.
  }\label{fig:R2_oriented}
\end{figure}

In Fig.~\ref{fig:R2_oriented}, the mean-square separation is shown for a range
of initial wall-normal positions $y_0$, separation distances $D_0$, and
separation orientations $\vb{e}_0$.
As can be predicted from the expression for the short-time regime
(Eq.~\ref{eq:R2_short_time_t3}) and the structure functions described in
Section~\ref{sec:structure_functions}, initial orientation plays an
important role for particles initialised near the wall (subfigures a-b), while
its impact is weaker far from the wall (subfigures c-d).
In all cases, the initial ballistic separation is
more efficient when the initial separation $D_0$ is larger.
Consistently with the behaviour of the velocity structure functions presented in
Fig.~\ref{fig:structure_functions}, anisotropy at
short times enhances separation when particles are initially oriented in the
wall-normal direction.
This has already been observed by \citet{Shen1997} in the case of homogeneous
turbulent shear flow.
The authors found that particles separate faster when they are initially
oriented in the cross-stream direction.
Furthermore, particles that are oriented in the spanwise direction separate
faster than those oriented in the streamwise direction.
As discussed in Section~\ref{sec:structure_functions}, the presence of streaks
and quasi-streamwise vortices near the wall \citep{Robinson1991}, implies weaker
velocity increments between two points aligned in the streamwise direction than
in the spanwise direction.

At very long times, the mean-square separation no longer depends on the initial
configuration of the pairs.
The curves from all the initial configurations collapse due
to loss of memory of the initial condition.
An intermediate time range connects the initial ballistic regime, strongly
dependent on the initial configuration, and the long-time dispersion regime,
independent of the initial configuration.
The ballistic time scale $t_0$ (represented by squares over each curve in
Fig.~\ref{fig:R2_oriented}) is an adequate time scale for representing the
transition from the ballistic
regime to the intermediate regime.
This regime is given by a super-diffusive process which is more efficient than the
initial ballistic regime, as already observed in
Section~\ref{sec:normalised_R2} (for pairs conditioned to an initial separation
$\abs{\DoVec} < 16\eta$).
From Fig.~\ref{fig:R2_oriented}, it is found that the slope of the
super-diffusive regime is steeper when the initial ballistic regime is slower,
that is, when the structure function $S_2(y_0, \DoVec)$ is weaker.
This is the case for smaller separations $\abs{\DoVec}$, and for wall-parallel
orientations, when the contribution of mean shear to the structure
function $S_2$ is zero.

\section{Mean shear influence}%
\label{sec:mean_shear}

In order to characterise the influence of mean shear on relative
dispersion in the channel, we decompose the time evolution of the
particle pair separation
into a separation induced by the mean velocity field,
$\overline{\vb{R}}(t)$, and a separation due to the fluctuating velocity field,
$\vb{R}'(t)$.
We then study the evolution of the mean-square separation resulting from the
fluctuating field $\RsqtPrime$.

\subsection{Decomposition of the mean-square separation}

Given an Eulerian mean velocity field $\vb{U}(\vb{x})$, we define the
fluctuating velocity of a fluid particle with trajectory $\vb{x}(t)$ as
$\vb{v}'(t) = \vb{v}(t) - \vb{U}(\vb{x}(t))$, where
$\vb{v}(t) = \vb{u}(\vb{x}(t), t)$ is the total fluid particle velocity.
Note that $\vb{v}'(t)$ is the fluctuating velocity field at the
particle position, $\vb{u}'(\vb{x}(t), t)$.
Correspondingly, a fluctuating acceleration can be defined as
\begin{equation}
  \vb{a}'(t)
  = \dv{\vb{v}'(t)}{t}
  = \vb{a}(t) - \vb{v}(t) \cdot \grad{\vb{U}(\vb{x}(t))},
  \label{eq:aprime}
\end{equation}
where $\dv*{}{t}$ is the Lagrangian derivative along the fluid particle path,
and $\vb{a}(t)$ is the total particle acceleration.
In channel flow, the mean velocity field takes the form
$\vb{U}(\vb{x}) = U(y) \, \vb{e}_x$, and therefore Eq.~\eqref{eq:aprime} writes
\begin{equation}
  \vb{a}'(t) = \vb{a}(t) - v_y(t) \dv{U(y(t))}{y} \vb{e}_x,
\end{equation}
where $y(t)$ and $v_y(t)$ are the wall-normal position and velocity of the
particle, respectively.

The increment of instantaneous separation between two particles $\vb{R}(t)$ and
their relative velocity
$\delta\vb{v}(t) = \vb{v}_2(t) - \vb{v}_1(t)$ are then linked by
\begin{align}
  \vb{R}(t)
  = \vb{D}(t) - \DoVec
  &= \int_0^t \delta\vb{v}(\tau) \dd{\tau} \\
  &= {\int_0^t \delta\vb{U}(\tau) \dd{\tau}}
  +  \int_0^t \delta\vb{v}'(\tau) \dd{\tau} \\
  &= \overline{\vb{R}}(t) + \vb{R}'(t),
\end{align}
where
$\delta\vb{U}(t) = \vb{U}(\vb{x}_2(t)) - \vb{U}(\vb{x}_1(t))$ is the
mean velocity field difference between the positions $\vb{x}_1(t)$ and
$\vb{x}_2(t)$ of the two particles, such that
$\vb{D}(t) = \vb{x}_2(t) - \vb{x}_1(t)$,
and $\delta\vb{v}'(t) = \vb{v}'_2(t) - \vb{v}'_1(t)$ is their relative
fluctuating velocity.

\begin{figure}[tb]
  \centering
  \includegraphics[]{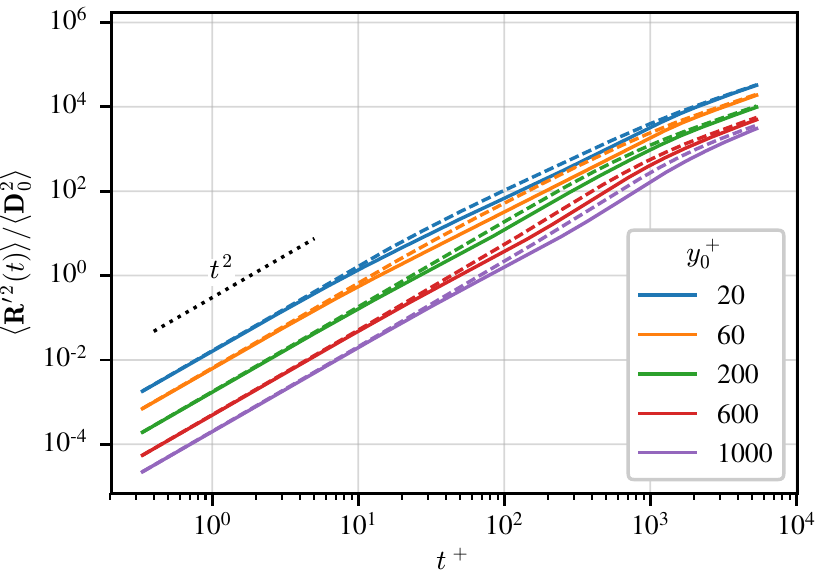}
  \caption{%
    Backward and forward mean-square separation due to the fluctuating flow $\RsqPrime$,
    normalised by the initial mean-square separation $\Lmean{\DoVec^2}$.
    Particle pairs are initially separated by $D_0 < 16\eta$ (dataset DS1).
    Different colours correspond to different initial wall distances $y_0^+$.
    Solid lines: forward dispersion.
    Dashed lines: backward dispersion.
  }\label{fig:R2prime_D0}
\end{figure}

The time evolution of $\RsqPrime$ is plotted in Fig.~\ref{fig:R2prime_D0} for
pairs initially separated by $\abs{\DoVec} < 16\eta$ and for different initial
wall-distances $y_0^+$.
By comparison to the $\Rsq$ shown in Fig.~\ref{fig:R2_D0}, the
mean-square separation induced by the fluctuating flow is about one order of
magnitude weaker than the total mean-square separation at very long times ($t^+
\approx \num{5000}$).
When the influence of mean shear is removed, the super-diffusive regime at
intermediate times is considerably weaker.
Regarding the initial ballistic regime, the difference between $\Rsq$ and
$\RsqPrime$ is more pronounced when pairs are initialised close to the wall.
For $y_0^+ = 20$, $\Rsq$ evolves considerably faster than $\RsqPrime$ during
the ballistic regime (that is, $\Svv > \SvvPrime$), as a result of the
dominant role of mean shear in the near-wall region.
For larger values of $y_0^+$, the influence of mean shear on the ballistic
regime is much weaker, implying that away from the wall the initial separation
regime (and thus the structure function $S_2$) is
governed by turbulent fluctuations.

Similarly to the total relative dispersion described in previous sections,
relative dispersion induced by the fluctuating flow is a time-asymmetric
process, with backward dispersion being faster than forward dispersion.
As before, this asymmetry is first evidenced as a deviation from the initial
ballistic separation.
The gap between backward and forward dispersion increases at intermediate times,
and then decreases at very long times.
This confirms that the temporal asymmetry of relative dispersion in turbulent channel flow
is a consequence of the irreversibility of turbulent fluctuations, as is
in isotropic flows \citep{Jucha2014}.

\subsection{Short-time dispersion}

Similarly to Eq.~\eqref{eq:R2_short_time_t3}, the short-time evolution of
the mean-square separation due to the fluctuating flow can be written as
\begin{equation}
  \RsqPrime[](y_0, \DoVec, t) =
  \SvvPrime t^2 + \SvaPrime t^3 + \order*{t^4}
  \quad
  \text{for }
  t \ll t_0',
  \label{eq:R2prime_short_time_t3}
\end{equation}
where $\DeltaV_0' = \DeltaV'(0)$ and $\DeltaA_0' = \DeltaA'(0)$.
Here $\DeltaA'(t) = \vb{a}_2'(t) - \vb{a}_1'(t)$ is the relative fluctuating
acceleration of the particles.
Thus, the separation $\vb{R}'(t)$ is also expected to follow an initial
ballistic growth, although the characteristic duration of this ballistic
regime is not necessarily the same as for the total change of separation
$\vb{R}(t)$.
From the above expression and according to the discussion in
Section~\ref{sec:ballistic_time_scales},
the ballistic time scale associated to $\vb{R}'$ is
defined as $t_0' = \SvvPrime / \abs*{\SvaPrime}$.

\begin{figure}[tb]
  \centering
  \includegraphics[]{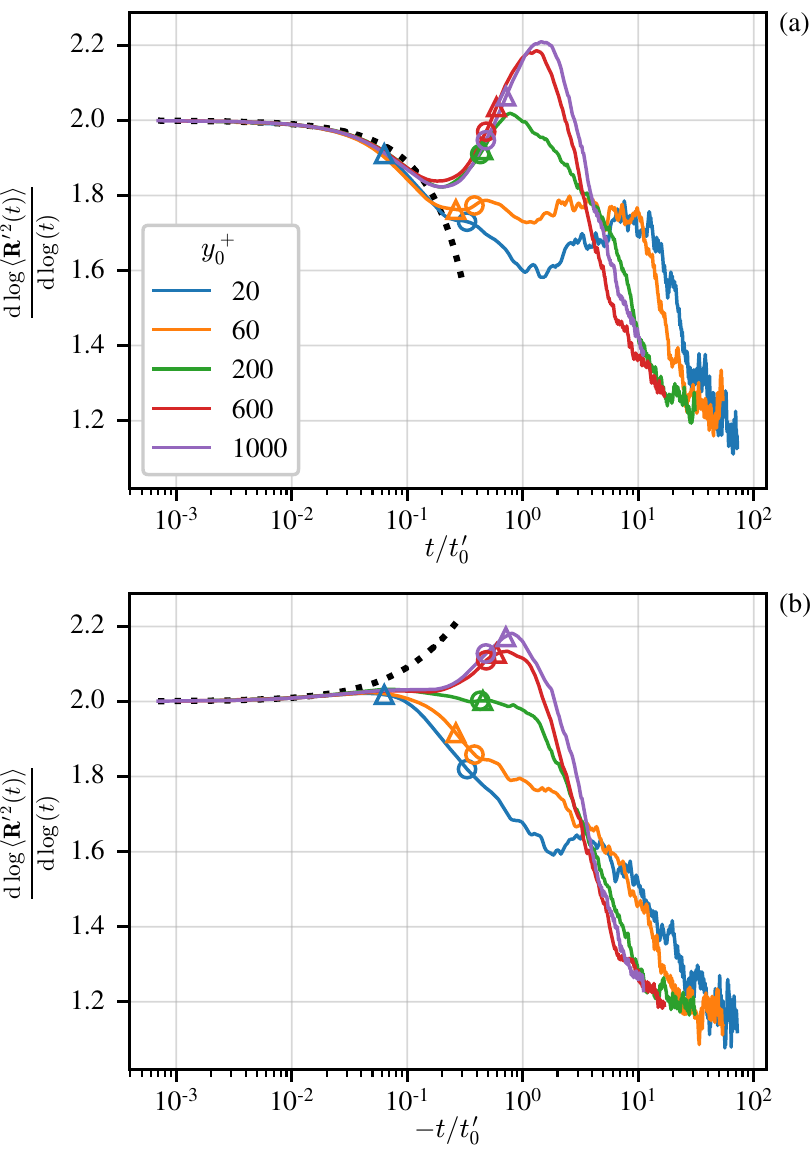}
  \caption{%
    Local scaling exponents of the mean-square separation by the fluctuating
    flow $\RsqtPrime$.
    (a) Forward dispersion.
    (b) Backward dispersion.
    The dotted line is derived from the truncated Taylor expansion
    $\RsqtPrime = \SvvPrime t^2 + \SvaPrime t^3$ with
    $\SvaPrime < 0$.
    For each initial wall distance $y_0^+$, markers indicate the local value of
    the Lagrangian integral time scale $T_L$ (circles) and of the mean shear
    time scale $T_S$ (triangles).
  }\label{fig:R2prime_logdiff}
\end{figure}

The local scaling exponent of $\RsqtPrime$ is shown in
Fig.~\ref{fig:R2prime_logdiff} with time normalised by $t_0'$.
As with $\Rsqt$ (shown in Fig.~\ref{fig:R2_D0}), for all wall distances the
initial ballistic regime is followed
by a decelerated separation in the forward case and by an accelerated separation
in the backward case, which are both explained by a negative value of
$\SvaPrime$ in Eq.~\eqref{eq:R2prime_short_time_t3}.
The observed behaviour closely follows the truncated Taylor expansion of
$\RsqtPrime$ at short times.
As it was observed from Fig.~\ref{fig:R2prime_D0}, the super-diffusive regime at
intermediate times is remarkably weaker for $\RsqtPrime$ than for the total separation
$\Rsqt$ (see Fig.~\ref{fig:R2_logdiff}), with maximum values that barely
exceed the initial ballistic scaling $\RsqPrime \sim t^2$.
This confirms that the intermediate super-diffusive regime that was found in
previous sections, described by an instantaneous scaling reaching
$\Rsq \sim t^3$, is due to mean shear and not to Richardson's law.
At very long times, the average separation rate decelerates continuously.
It may be predicted that the diffusion due to the fluctuating flow should tend
to a normally-diffusive process, as in HIT \citep{Taylor1922}, which
would correspond to a scaling $\RsqPrime \sim t$.
However, the available data is insufficient to verify this statement.

\begin{figure}[tb]
  \centering
  \includegraphics[]{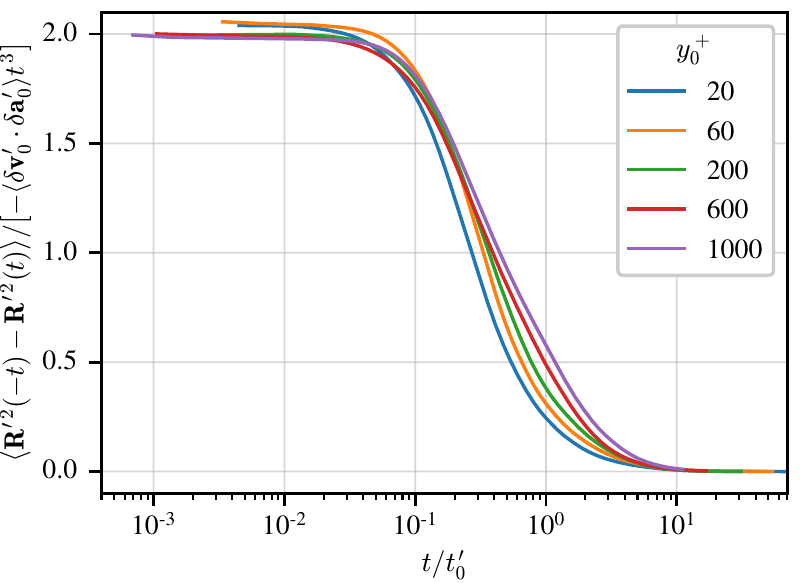}
  \caption{%
    Difference between backward and forward mean-square separation due to the
    fluctuating flow, compensated by $-\SvaPrime t^3$.
    Results were obtained from dataset DS1.
  }\label{fig:R2prime_time_asymmetry}
\end{figure}

Analogously to Eq.~\eqref{eq:R2_short_time_asymmetry}, the temporal asymmetry of
$\RsqtPrime$ can be described at short times by
\begin{equation}
  \RsqPrime[(t)] - \RsqPrime[(-t)] =
  2 \SvaPrime t^3 + \order*{t^5}
  \quad
  \text{for }
  t \ll t_0'.
  \label{eq:R2prime_short_time_asymmetry}
\end{equation}
The validity of this analytical prediction is verified from simulation data in
Fig.~\ref{fig:R2prime_time_asymmetry}, where the difference
$\RsqPrime[(t)] - \RsqPrime[(-t)]$ is plotted compensated by $\SvaPrime t^3$.
The expected plateau at $2$ is recovered for times $t \lesssim 0.1 t_0'$,
similarly to the case of the total mean-square separation $\Rsq$
(Fig.~\ref{fig:R2_time_asymmetry}), and consistently with equivalent results in
HIT \citep{Jucha2014, bragg_forward_2016}.
Namely, in the case of HIT, \citet{Jucha2014} compared the compensated
difference as given in Fig.~\ref{fig:R2prime_time_asymmetry} obtained from
DNS with experimental data at four different Reynolds numbers ranging from
$\Rey_\lambda =$ 200 to 690.
All their data showed a clear plateau up to $t \approx t_0 / 10$, in complete
agreement with equation~\eqref{eq:R2prime_short_time_asymmetry}.
Here, the plateau ranges up to $t \approx t_0' / 20$ for different wall
distances and even close to the wall, where anisotropy increases.
Interestingly, for $y_0^+ = 20$, the plateau is present and the curve matches
the behaviour at higher wall distances, which was not the case for $\Rsq$ shown
in Fig.~\ref{fig:R2_time_asymmetry}.
More generally, the spread of the curves associated to different $y_0^+$ is
reduced with respect to that obtained from the total mean-square separation
$\Rsq$ (Fig.~\ref{fig:R2_time_asymmetry}), emphasising the impact of mean shear
on $\Rsq$, at relatively short times close to the wall and at larger times away
from the wall.
As mentioned in Section~\ref{sec:numerical_approach}, the initial wall distance
of particles in the $y^+_0 = 20$ set is within $0 \lesssim y^+ \lesssim 40$.
Therefore, the current study does not allow for a finer description of
the temporal asymmetry of pair dispersion close to the wall.

\begin{figure}[tb]
  \centering
  \includegraphics[]{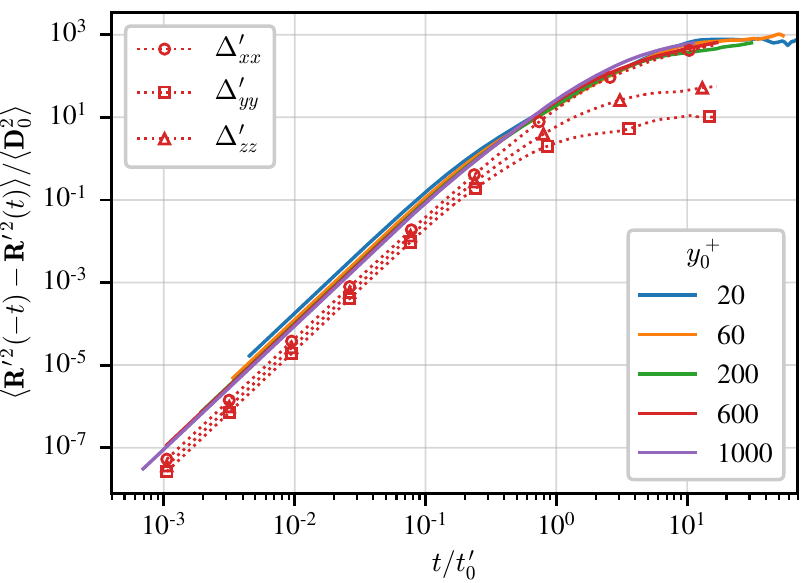}
  \caption{%
    Difference between forward and backward mean-square separation due to the
    fluctuating flow, compensated by $\Lmean{\DoVec^2}$.
    Results were obtained from dataset DS1.
    Dotted lines represent the directional decomposition associated to
    dispersion tensor components $\Dij[xx]'$ (circles), $\Dij[yy]'$ (squares)
    and $\Dij[zz]'$ (triangles), for pairs in the $y_0^+ = 600$ set.
  }\label{fig:R2prime_D0sq_time_asymmetry_xyz}
\end{figure}

Figure~\ref{fig:R2prime_D0sq_time_asymmetry_xyz} plots the difference
$\RsqPrime[(-t)] - \RsqPrime[(t)]$ compensated by the initial mean-square
separation $\Lmean{\DoVec^2}$.
As in Fig.~\ref{fig:R2_time_asymmetry}, all the curves display a positive sign
associated to backward dispersion being faster than forward dispersion.
Similarity of the results is found for all initial wall distances including
$y_0^+ = 20$.
Moreover, a common long-time limit is observed.
This limit is characterised by a plateau starting at $t \approx 10 t_0'$,
suggesting that the temporal asymmetry of dispersion is enhanced during the
short-time separation regime, and then becomes negligible at long times.
The dotted lines in Fig.~\ref{fig:R2prime_D0sq_time_asymmetry_xyz} represent the
forward-backward dispersion difference in each Cartesian direction (i.e.\ the
contribution of each separation component to
Eq.~\eqref{eq:R2prime_short_time_asymmetry}), for particles initially located at
$y_0^+ \approx 600$.
In this case, backward dispersion is faster than forward dispersion in every
direction.
At long times, the time asymmetry of the dispersion is most pronounced in the
streamwise direction.

In the case of a mean shear turbulent flow, \citet{Celani2005} estimated the
time required for two particles to reach separations at which the mean shear and
the turbulent fluctuations contributions become comparable.
According to the authors, this time scale $t_c$ is inversely proportional to the
mean shear, and therefore directly proportional to $T_S = {(\dd U(y) / \dd y)}^{-1}$.
In turbulent channel flow, as $y^+$ increases, mean shear decreases and the time
scale at which mean shear and turbulent fluctuations present comparable
contributions increases.
This is confirmed by the results in Fig.~\ref{fig:R2_oriented} since the squares
on the curves representing the ballistic time scale $t_0$ move to the right as
the wall distance increases.

\section{Relative dispersion tensor}%
\label{sec:dispersion_tensor}

Until now, we have considered statistics related to the change of
separation magnitude between a pair of particles,
$\abs{\vb{R}(t)} = \abs{\vb{D}(t) - \DoVec}$.
However, the separation between two particles in inhomogeneous and anisotropic
flows presents an anisotropic evolution in time.
Namely, the presence of mean shear enhances particle
separation in the streamwise direction, while it does not have a
direct effect in the other directions.

The anisotropy of relative dispersion can be investigated by means of the
relative dispersion tensor \citep{Batchelor1952, Monin1975},
\begin{equation}
  \Dij(t)
  = \Lmean{R_i(t) R_j(t)},
\end{equation}
where $R_i(t) = D_i(t) - D_i(0)$ is the $i$-th component of $\vb{R}(t)$.
The trace of $\Dij$ is equal to the mean-square separation,
$\Dij[ii](t) = \Rsqt$.
By construction, $\Dij$ is a symmetric tensor.
In channel flow, due to the statistical symmetry $z \leftrightarrow -z$, its
non-diagonal components $\Dij[xz]$ and $\Dij[yz]$ are zero.
As a consequence, the relative dispersion tensor contains a single independent
non-diagonal component, $\Dij[xy] = \Dij[yx]$.
Each component of the relative dispersion tensor depends on the initial
wall distance $y_0$ and on the initial particle separation vector $\DoVec$.

The short-time evolution of $\Rsqt$ as predicted by
Eq.~\eqref{eq:R2_short_time_t3} can be generalised to
\begin{multline}
  \Dij(y_0, \DoVec, t)
  = \Lmean{\delta v_{0i} \delta v_{0j}} t^2
  +
  \left(
    \Lmean{\delta v_{0i} \delta a_{0j}}
    +
    \Lmean{\delta v_{0j} \delta a_{0i}}
  \right) \frac{t^3}{2}
  \\
  + \order*{t^4}
  \quad \text{for } t \ll t_0.
  \label{eq:Dij_short_time}
\end{multline}
Therefore, each component of $\Dij$ independently follows an initial ballistic
regime according to the velocity structure function tensor
$\Lmean{\delta v_{0i} \delta v_{0j}}
\allowbreak = S_{ij}(\vb{x}_0, \DoVec)
\allowbreak = \overline{
  \delta u_i(\vb{x}_0, \DoVec) \,
  \delta u_j(\vb{x}_0, \DoVec)
}$.
At the next order, the $t^3$ term is governed by the symmetric part of the
crossed velocity-acceleration structure function tensor
$\Lmean{\delta v_{0i} \delta a_{0j}}
= \overline{
  \delta u_i(\vb{x}_0, \DoVec) \,
  \delta a_j(\vb{x}_0, \DoVec)
}$.

Due to wall confinement, particle separation in the wall-normal
direction cannot exceed $\abs*{D_y} = 2h$.
It is possible to estimate, at sufficiently long times, the
influence of confinement on the wall-normal mean-square separation $\Lmean{D_y^2}$.
Under the assumption of loss of memory of the initial particle position, the
wall-normal position of a single particle can be expected to follow a uniform
distribution at long times, described by the probability density function (PDF)
$P_y(y) = 1/(2h)$ for $0 \leq y \leq 2h$.
Moreover, the trajectories of two particles in a pair are expected to
decorrelate over a sufficiently long time, implying that the
joint PDF describing the wall-normal positions of the two particles,
$P_{yy}(y_1, y_2)$, writes as $P_y(y_1) P_y(y_2)$.
Under these assumptions, the wall-normal mean-square
separation is given by
\begin{equation}
  \Lmean{D_y^2}
  \equiv
  \int\limits_0^{2h} \int\limits_0^{2h}
  {(y_2 - y_1)}^2 P_{yy}(y_1, y_2) \dd{y_1} \dd{y_2}
  = \frac{2}{3} h^2.
\end{equation}
This equation is also an estimation for the wall-normal component of the
dispersion tensor, i.e.\ $\Dyy \approx 2h^2/3$ at long times, under
the additional assumption that the initial wall-normal separation is
small compared to the channel dimensions, i.e.\ $|D_{0y}| \ll h$.

\subsection{Short-time dispersion}

\begin{figure}[tp]
  \centering
  \includegraphics[]{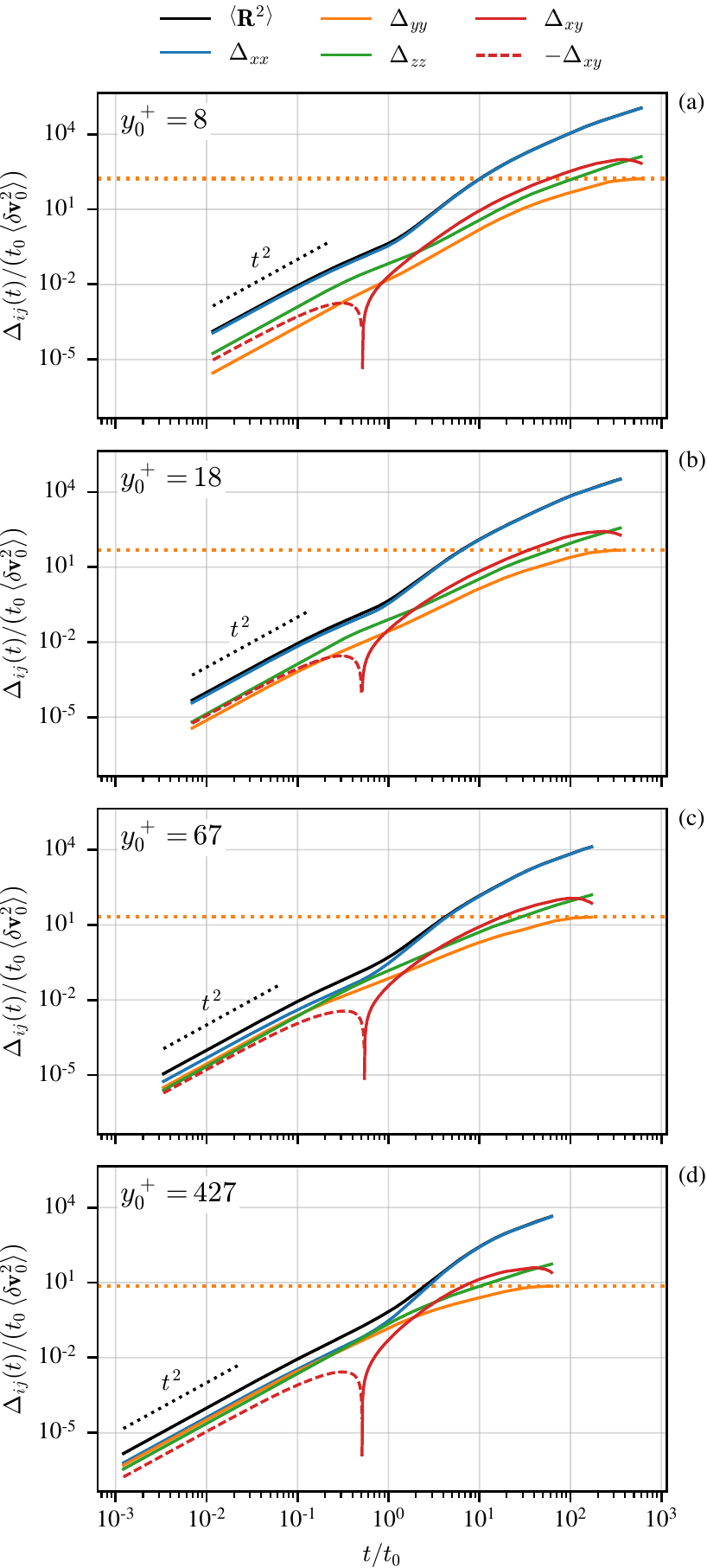}
  \caption{%
    Components of the relative dispersion tensor, normalised by the structure
    function $\Svv$ and the characteristic ballistic time $t_0$.
    Particle pairs are initially located at $y_0^+ =$ 8, 18, 67 and 427
    (subfigures (a) to (d)).
    Pairs are initially oriented in the spanwise direction ($z$), with a
    separation $D_0 = 16\eta$.
    The total mean-square separation $\Dij[ii] = \Rsq$ is also represented.
    The dotted horizontal line marks the level $\Dij = 2h^2/3$, where $h$ is the
    channel half-width.
    Results were obtained from dataset DS2.
  }\label{fig:Dij_D16_z}
\end{figure}

In Fig.~\ref{fig:Dij_D16_z}, the temporal evolution of the relative dispersion
tensor is shown for particle pairs initially located at different wall-distances
$y_0$.
In all cases, pairs are initially separated in the spanwise direction by
$D_0 = 16\eta$.
Because of the spanwise alignment of the pairs, mean shear does not play a role
during the initial ballistic separation.
As predicted by Eq.~\eqref{eq:Dij_short_time}, the ballistic regime is observed
for each component of $\Dij$.
Pair dispersion is anisotropic since the start of the separation.
During the ballistic regime, for the shown initial configurations, particles
near the wall separate faster in the streamwise direction, while separation is
slowest in the wall-normal direction.
For the nearest wall distances ($y_0^+ =$ 8 and 18), this means that the
streamwise separation dominates the total separation from the start, which is
confirmed by the superposition between the curves for $\Dxx$ and $\Rsq$ at
all times.
By DNS in a homogeneous turbulent shear flow, \citet{Shen1997} also found that
particle-pair dispersion is most effective in the streamwise direction, as
already stated in the introduction.
In turbulent channel flow, the rapid streamwise separation at short times for
particles initially separated in the spanwise direction may be explained by the
presence
of near-wall streaks.
These are elongated regions in the streamwise direction, carrying low-speed and
high-speed fluid alternating in the spanwise direction \citep{Robinson1991}.
Two particles initially belonging to two neighbouring streaks (a high-speed
streak next to a low-speed streak), experience a rapid streamwise
separation due to the velocity difference between the streaks.

As expected, the short-time behaviour approaches isotropy as particles are
released further away from the wall.
In all cases, for each of the three diagonal components, the ballistic
separation is immediately followed by a deceleration of the separation rate.
Following Eq.~\eqref{eq:Dij_short_time}, and consistently with the observations
from previous sections, this deceleration is associated with a negative value of the
component-wise crossed structure functions $\Lmean{\delta v_{0i} \delta a_{0i}}$
(where repeated indices do not imply summation).

\subsection{Intermediate and long-time dispersion}

At intermediate times starting from $t \approx t_0$, $\Dxx$ displays an
accelerated separation rate, while $\Dyy$ and $\Dzz$ evolve at slower rates
compared to the initial ballistic regime.
The rapid separation in the streamwise direction can be attributed to the effect
of mean shear.
The duration of this rapid separation regime, which lasts until
$t \approx 10 t_0$, is consistent with the duration of the super-diffusive
regime observed for $\Rsq$ in Fig.~\ref{fig:R2_logdiff}a and discussed in
Section~\ref{sec:mean_shear}.

The estimation $\Dyy \approx 2h^2/3$ accurately predicts the
wall-normal pair separation at long times.
This prediction is valid starting from $t^+ \approx 10^4$ for all initial
wall distances $y_0^+$ (not shown here).
As noted in Section~\ref{sec:initial_separation}, Fig.~\ref{fig:R2_oriented},
for $t^+ \approx 10^4$ the mean-square separation no longer
depends on the initial pair configuration.

\subsection{Time evolution of cross-term \texorpdfstring{$\Dxy$}{Dxy}}%
\label{sec:dispersion_tensor_Dxy}

Finally, the time evolution of the cross-term $\Dxy$ may yield additional
insight on the mechanisms of pair separation in wall-bounded turbulence.
Initially, $\Dxy$ evolves ballistically with an increasingly negative value at
all wall distances, which following Eq.~\eqref{eq:Dij_short_time} corresponds to
a negative value of the structure function
$\Lmean{\delta v_{0x} \delta v_{0y}}$.
This is consistent with the model of \citet{Yang2017}, predicting a
structure function $S_{xy}^+$ between \num{-1} and \num{-2} when $y_0$ and
$y_0 + D_{0y}$ are within the logarithmic region.
The ballistic regime ends with a deviation of $\Dxy$ towards positive values,
resulting from a positive value of the $t^3$ term in
Eq.~\eqref{eq:Dij_short_time}.
This leads to a change of sign of $\Dxy$, that becomes positive at $t \approx
t_0/2$ for all initial wall distances.

At intermediate times, $\Dxy$ displays a rapid growth, coinciding with the
super-diffusive growth of $\Dxx$.
As for $\Dxx$, this is due to the influence of mean shear.
To illustrate this, we consider a pair of particles A and B initially located
in the lower half
of the channel ($0 < y_0 < h$).
At some point, even if the particles are initially close, their wall-normal
separation $|D_y| = |y_B - y_A|$ will grow due to turbulent diffusion
until $|D_y|$ becomes large enough for mean shear effects to be important.
Without loss of generality, we assume that particle B is further away from the
wall than particle A, i.e.\ $D_y > 0$.
Therefore, as long as the particles have not yet crossed the channel centre,
particle B is located in a region of faster average flow than A, and thus their
streamwise separation $D_x = x_B - x_A$ grows rapidly due to the mean shear.
The result is a product $D_x D_y$ which rapidly grows over time as long as $D_y$
remains positive.
This is no longer valid once a particle crosses the channel
centre, leading to the decelerated growth of $\Dxy$ at later times.

\section{Ballistic dispersion model}%
\label{sec:ballistic_dispersion_model}

Relative dispersion statistics in turbulent channel flow may be reproduced using a
simple model based on the ballistic cascade phenomenology proposed by
\citet{Bourgoin2015} to describe relative dispersion in isotropic turbulent
flows.
\citet{Bourgoin2015} explained the transition from the short-term ballistic
separation to Richardson's super-diffusive regime ($\Rsq \sim t^3$) at long
times as a temporal progression of discrete, short-lived ballistic separations.
This approach is similar to previous models
\citep{Sokolov2000, Faber2009, Thalabard2014}, all of which considered the
relevance of successive ballistic separations on pair dispersion.
In the following, we briefly present Bourgoin's ballistic cascade model in the
case of 3D isotropic turbulent flows.
Then, we propose and test a modified model that takes into account mean shear in
the case of turbulent channel flows.

\subsection{Ballistic cascade model in isotropic turbulence}

\citet{Bourgoin2015} formulated the ballistic cascade in isotropic turbulence as
a simple iterative model.
Starting from an initial separation $D_0$ within the inertial subrange, the
mean-square separation $\Dsq$ is incremented at each iteration by a ballistic
assumption according to:
\begin{equation}
  D_{k+1}^2
  = D_k^2 + S_2(D_k) \, t_k^{\prime 2}(D_k)
  \quad
  \text{for } k = 0, 1, 2, \ldots,
  \label{eq:model_HIT}
\end{equation}
where $D_k^2$ is the mean-square separation at iteration $k$.
Here, $S_2(D_k) = \frac{11}{3} C_2 {(\varepsilon D_k)}^{2/3}$ is the isotropic
second-order Eulerian velocity structure function for $D_k$ in the inertial
subrange, as introduced in Section~\ref{sec:short_time_dispersion}.
The duration of the $k$-th iteration is given by $t_k' = \alpha t_k$, where
$t_k = S_2(D_k) / (2 \varepsilon)$ is a characteristic time of the ballistic
regime (equal to $t_I$ as defined in
Section~\ref{sec:ballistic_time_scales}), and $\alpha$ is a non-dimensional
constant referred to as the persistence parameter.
The total time elapsed by the start of iteration $k$ is
$T_k = \sum_{j = 0}^{k - 1} t_j'(D_j)$.

Besides Kolmogorov's constant $C_2$, which has the well-accepted value $C_2
\approx {2.1}$ \citep{Sreenivasan1995}, $\alpha$ is the only free parameter of
the model.
By analytically relating $C_2$ and $\alpha$ to Richardson's constant,
\citet{Bourgoin2015} found $\alpha = 0.12$ as the value that best matches the
well-accepted Richardson constant in 3D turbulence,
$g \approx 0.55$ \citep{ott_experimental_2000, Bitane2012}.
With this value of the persistence parameter, the ballistic cascade
model has been shown to reproduce with great accuracy the DNS results from
\citet{Bitane2012} in HIT at a Taylor-scale Reynolds number $\Rey_\lambda =
730$, with initial particle separations $D_0$ ranging between $2\eta$ and
$48\eta$.

The model described by Eq.~\eqref{eq:model_HIT} is symmetric in time.
\citet{Bourgoin2015} also proposed a time-asymmetric version of the model by
taking into account the $t^3$ term in the Taylor
expansion~\eqref{eq:R2_short_time_t3}, associated with the velocity-acceleration
structure function $\Sau$.
This refined model captures a ratio between backward and forward
Richardson constant $g_{\text{bw}}/g_{\text{fw}} = 1.9$, consistent
with available experimental and DNS results.

\subsection{Ballistic cascade model in inhomogeneous turbulence}%
\label{sec:ballistic_model_inhomogeneous}

\newcommand*{\xtilde}{\tilde{\vb{x}}}
\newcommand*{\Dvec}{\vb{D}}
\newcommand*{\xkDk}{\xtilde_k, \Dvec_k}

As shown in previous sections, the mean-square separation of particle pairs in
channel flow is accurately described at short times by an average
ballistic separation.
Therefore, a model based on a succession of ballistic separations may seem suitable
for predicting pair dispersion statistics in the studied flow.
In the following, such a model is proposed based on Bourgoin's approach, which
is adapted to account for the effect of an
inhomogeneous mean velocity field $\vb{U}(\vb{x})$.
The model is also adjusted to take into account the transition from inertial to
integral-scale separations at sufficiently long times.
In addition to the mean velocity field, the present model requires as input the
mean turbulent dissipation rate $\varepsilon(\vb{x})$.
The model is started with an initial pair separation vector $\DoVec$ and with
the initial position of the pair centroid, $\xtilde =
(\vb{x}_0^A + \vb{x}_0^B) / 2$, where $\vb{x}_0^A$ and $\vb{x}_0^B$ are the
initial positions of the two particles.
In channel flow, due to homogeneity in the streamwise and spanwise directions,
the model requirements reduce to the mean streamwise velocity profile along the
channel $U(y)$ and the turbulent dissipation profile $\varepsilon(y)$, as well
as the initial particle configuration given by $\DoVec$ and the wall-normal centroid
position $\tilde{y}_0 = (y_0^A + y_0^B) / 2$.

We model the time evolution of the mean-square separation vector $\Dsq$ and the
position of the pair centroid $\xtilde$ iteratively.
As a first approximation, the centroid position is kept fixed over time,
i.e.\ $\xtilde_k = \xtilde_0$ at every iteration $k$.
This will be improved in future versions of the model, by taking into account
the drift of the particle pair centroid based on single-particle dispersion
statistics.
At iteration $k$, the mean-square separation in each direction
$i \in \{x, y, z\}$ is incremented according to
\begin{equation}
  D_{k+1, i}^2
  = D_{k, i}^2 + S_{2i}(\xkDk) \,t_k^{\prime 2}(\xkDk)
  \quad
  \text{for } k = 0, 1, 2, \ldots,
  \label{eq:model_inhomog}
\end{equation}
where $D_{k, i}^2$ is the mean-square separation in the $i$-th direction at
iteration $k$.
The total mean-square separation is then
$D_k^2 = D_{k, x}^2 + D_{k, y}^2 + D_{k, z}^2$.
The structure function $S_{2i}$ is associated to the velocity component $u_i$,
and can be written as the superposition of a mean and a fluctuating component:
\begin{equation}
  S_{2i}(\xkDk)
  = \overline{S}_{2i}(\xkDk) + S'_{2i}(\xkDk).
\end{equation}
The mean component is readily obtained from the mean velocity field:
\begin{equation}
  \overline{S}_{2i}(\xkDk)
  =
  {\left[
      U_i\left(\xtilde_k + \frac{\Dvec_k}{2}\right) -
      U_i\left(\xtilde_k - \frac{\Dvec_k}{2}\right)
  \right]}^2.
  \label{eq:model_S2i_mean}
\end{equation}
The fluctuating component $S_{2i}'$ is estimated so as to account for the
transition from inertial to integral-scale separations.
For separations within the inertial range, $S_{2i}'$ is estimated from HIT as
\citep{Pope2000}:
\begin{equation}
  S_{2i}^I(\xkDk)
  =
  C_2 {\Big( \varepsilon(\xtilde_k) \abs{\Dvec_k} \Big)}^{2/3}
  \left(
    \frac{4}{3} - \frac{1}{3} \frac{D_{k, i}^2}{\abs{\Dvec_k}^2}
  \right).
\end{equation}
In HIT, the structure function $S_2(D_0)$ tends to $2\sigma_u^2$ for
integral-scale separations, over which the velocity field becomes fully
decorrelated in space.
Here $\sigma_u^2$ is the variance of the velocity fluctuations.
Consistently, the present model estimates $S_{2i}'$ for separations within the
integral scales as
\begin{equation}
  S_{2i}^L(\xkDk) = 2\sigma_i^2(\xtilde_k),
\end{equation}
where $\sigma_i^2 = \overline{u_i^{\prime 2}}$ is the variance of the velocity
component $u_i$.
It is reasonable to model $S_{2i}'$ as an increasing function of the spatial
increment $\abs*{D_{k,i}}$.
Therefore, a straightforward way of estimating the fluctuating component of the
velocity structure function is to take
\begin{equation}
  S'_{2i}(\xkDk) = \min \left\{ S_{2i}^I(\xkDk), S_{2i}^L(\xkDk) \right\}.
\end{equation}
According to this expression, the transition from inertial to integral
separations is implicit, since it happens once the inertial-range prediction
$S_{2i}^I$ becomes larger than $S_{2i}^L$.
A weakness of this model is that the scale transition happens abruptly, whereas
the structure function should be a smooth function of the separation.
According to the present model, the three components of
the separation $\vb{D}$ may transition to the integral scales at different
times.
This is not a problem since, in inhomogeneous flows, the integral scales
generally depend on the considered orientation.

As in the original model by \citet{Bourgoin2015}, the iteration time is
taken as $t_k' = \alpha t_k$, with the ballistic time scale estimated as
$t_k(\xkDk) = S_2'(\xkDk) / (2 \varepsilon(\xtilde_k))$.
Here, the structure function $S_2'$ is given by
$S_2'(\xkDk) = \sum_{i = 1}^3 S_{2i}'(\xkDk)$.
The value of the persistence parameter $\alpha = 0.12$ is kept unchanged.
Finally, the time elapsed by the start of iteration $k$ is given by
$T_k = \sum_{j = 0}^{k - 1} t_j'(\xtilde_j, \Dvec_j)$.
When the mean velocity field is constant, mean shear is neglected and the
present model falls back to the isotropic model described above when separations
are in the inertial subrange.

The present model does not account for the presence of solid boundaries.
In the case of channel flow, this implies that wall confinement is not accounted
for.
Hence, the present model allows particles to travel beyond the channel walls.
As implied by Eq.~\eqref{eq:model_S2i_mean}, the model effectively estimates the
absolute wall-normal position of the two particles as
$y_k = \tilde{y}_k \pm D_{k,y}/2$, where $\tilde{y}_k$ is the wall-normal
position of the pair centroid.
When one of the particles crosses the channel walls, its mean velocity is
taken as $\vb{U} = 0$.
In future work wall confinement will be accounted for thoroughly.
We present here preliminary findings of a very simple extension of the original
ballistic model in HIT\@.

The proposed simple model is tested in the channel flow configuration, using as input a
mean velocity profile $U(y)$, a turbulent dissipation profile $\varepsilon(y)$,
and velocity variance profiles $\overline{u_i^{\prime 2}}(y)$
obtained from our DNS at $\Rey_\tau = \num{1440}$.
We test two initial configurations, corresponding to initial particle locations
$y_0^+ =$ 67 and 427.
In both cases, the initial particle separation is
$\DoVec = 16 \eta \vb{e}_z$.
The DNS results corresponding to these cases were already analysed in the
previous sections (see for instance Fig.~\ref{fig:Dij_D16_z}).
Since the present model includes elements from isotropic turbulence, its results
are expected to be more accurate for particles initialised far from the walls,
where anisotropy is weaker.
The chosen initial separation $D_0 = 16\eta$ is rather favourable for testing
the model since, as shown in Fig.~\ref{fig:structure_functions}, the structure
function $S_2$ closely matches the expected inertial-range behaviour from HIT
for this initial separation.
For smaller separations such as $D_0/\eta =$ 1 and 4, the model should be
extended by including the dissipation-range structure functions as estimated in
Section~\ref{sec:structure_functions}.

\begin{figure}[tb]
  \centering
  \includegraphics[]{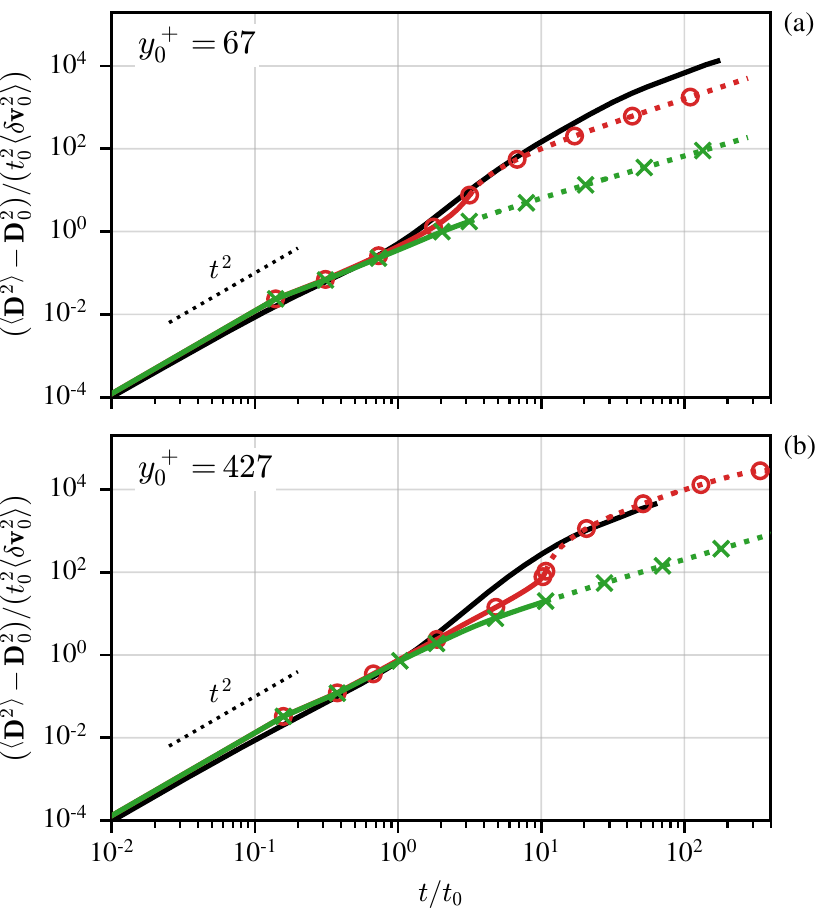}
  \caption{%
    Inhomogeneous ballistic cascade model compared to channel flow DNS results.
    Particle pairs are initially located at (a) $y_0^+ = 67$ and (b) $y_0^+ =
    427$.
    In both cases, the initial pair separation is $D_0 = 16\eta$ in the spanwise
    direction.
    Black line, DNS results;
    circles, ballistic model with velocity profile $U(y)$;
    crosses, ballistic model with constant velocity profile.
    The dotted part of the model curves correspond to the results once
    one of the particles has traversed the channel walls.
  }\label{fig:model_channel}
\end{figure}

A comparison between the model and the DNS results is shown in
Fig.~\ref{fig:model_channel} for the two chosen initial configurations.
Also shown is a variant of the model with a zero mean velocity profile
($U(y) = 0$).
As mentioned above, this is equivalent to neglecting the effect of mean shear on
dispersion.
Furthermore, since the mean turbulent dissipation $\varepsilon$ does not vary in
time in the
inhomogeneous model (because the particle pair centroid $\xtilde$ does not
move), the model variant is actually equal to the homogeneous model by
\citet{Bourgoin2015} as long as the separations $D_{k,i}$ stay within the
inertial subrange.

As shown in the Fig.~\ref{fig:model_channel}, during the first few ballistic
iterations both versions give quite satisfactory predictions compared to the DNS
results.
Model predictions up to $t \approx 3
t_0$ closely follow the DNS results for the initial wall distance
$y_0^+ = 67$.
Up to $t \approx t_0$, the two versions of the model show a similar behaviour,
implying the absence of mean shear influence.
Still, the models closely predict a deceleration of pair separation after the
initial ballistic regime.
Moreover, the full model predicts the start of the super-diffusive regime that follows,
although it does not precisely capture the time at which this regime starts
being observed.
The model with zero velocity profile does not show evidence of Richardson's
$t^3$ regime due to a lack of scale separation, since particle pairs
do not spend enough time in the inertial subrange.

Ongoing work is dedicated to a more refined model that partially accounts for
wall confinement through the particle pair centroid trajectory, which is pushed
away from solid boundaries.
At long times, when the memory of the initial particle position is lost,
the particle pair centroid is expected to be located, in average, at the channel
centre.
The present model will also be extended to account for the inter-dependency
between separation directions.
This will be quantified by the non-diagonal components of the relative
dispersion tensor described in Section~\ref{sec:dispersion_tensor}.
The model described by Eq.~\eqref{eq:model_inhomog} will then be rewritten
according to a tensor formulation.
In channel flow, this requires the estimation of the crossed velocity structure
function
$S_{xy}(y, \vb{D})
= \overline{\delta u_x(\vb{x}, \vb{D}) \, \delta u_y(\vb{x}, \vb{D})}$.

\section{Conclusions}%
\label{sec:conclusion}

This work deals with forward and backward dispersion statistics of fluid particle
pairs in a turbulent channel flow obtained by direct numerical simulations.
Relative dispersion statistics are conditioned to a wide range of initial
configurations.
Each configuration is given by an initial separation, orientation, and wall
distance of an ensemble of particle pairs.

Irrespectively of the initial pair configuration, the
mean-square particle separation at short times is accurately
described by the Eulerian structure of the flow at the initial configuration,
namely by the second-order velocity structure function $S_2(y_0, \DoVec)$ and
the crossed velocity-acceleration structure function $\Sau(y_0, \DoVec)$.
The characteristic time scale derived from these two statistics has been shown
to represent the duration of the short-time regime.
The initial ballistic regime is typically driven by turbulent fluctuations.
However, when the initial wall-normal separation is larger
than the local characteristic shear length scale, the influence of mean shear on
the separation rate is evidenced.

The short-time evolution is followed by a shear-driven super-diffusive regime.
Its scaling is highly dependent on the
initial particle configuration.
Namely, the separation rate in this regime is most important in cases where mean shear does
not play a role at short times (for small initial separations, or for initial
orientations parallel to the wall).
Conversely, when shear affects the ballistic regime, the
intermediate regime presents lower separation rates.

Consistently with similar studies in isotropic turbulence, particle pairs
separate faster in average when followed backwards than forwards in time.
At short times, this time asymmetry is associated with a negative sign of
$\Sau$.
The asymmetric behaviour persists when only the separation by the fluctuating
flow is considered, confirming that the observed time asymmetry is a consequence
of turbulence irreversibility.

The anisotropy of relative dispersion has been characterised by studying the
relative dispersion tensor $\Dij$.
The dominant role of mean shear is described by an increased growing rate of the
streamwise mean-square separation $\Dxx$ and of the cross-term $\Dxy$ during the
intermediate regime that follows ballistic growth.
Conversely, the wall-normal and spanwise diagonal terms $\Dyy$ and $\Dzz$
display a decelerated growth following the ballistic regime.

Finally, a simple relative dispersion model has been introduced based on the
ballistic cascade phenomenology proposed by \citet{Bourgoin2015} for
isotropic flows.
The present model accounts for the effect of mean shear on pair separation,
while keeping strong isotropic assumptions of the original model.
When particle pairs are initialised away from the wall, with initial separations
within the inertial range, the model closely predicts the mean-square separation
obtained from channel flow DNS over short times.
Later, the model predicts the accelerated separation of particles due to mean
shear, although the quantitative comparison with the DNS data remains
unsatisfactory.
In future developments of the inhomogeneous model,
wall confinement effects on particle displacement will be accounted for.

One of the most prominent features of wall-bounded turbulent flows are near-wall
streamwise vortices, responsible for ejections and sweeps.
The role of these particular structures on the short-time dispersion of particle
pairs will be examined in future studies.
Forthcoming studies will also deal with Lagrangian dispersion of fluid
particle tetrads and analysis of four-point velocity difference statistics.

\section*{Acknowledgements}

This work has been supported by Agence Nationale de la Recherche (Grant
No.\ ANR-13-BS09-0009).
Simulations have been performed on the P2CHPD cluster of the Fédération
Lyonnaise de Sciences Numériques and the national computing centre CINES (grant
no.\ DARI A0022A07707).
J.I.P. is grateful for CONICYT Becas Chile Grant No.~72160511 for supporting his
work.

\section*{References}
\bibliography{biblio}

\end{document}